\documentclass[prd,reprint,onecolumn,superscriptaddress,tightenlines,nofootinbib,eqsecnum]{revtex4-2}

\usepackage{amsmath}
\usepackage{cancel}
\usepackage{amsfonts}
\usepackage{amssymb}
\usepackage{bm}
\usepackage[colorlinks]{hyperref}
\usepackage{mathrsfs}
\usepackage{graphicx}
\usepackage{multirow}
\usepackage{empheq}
\usepackage{ulem}
\usepackage{tensor}
\usepackage{tabularx}
\usepackage[usenames]{color}
\definecolor{darkgreen}{rgb}{0,0.6,0}
\hypersetup{urlcolor=darkgreen, citecolor=darkgreen}
\usepackage{cleveref}
\newcolumntype{Y}{>{\centering\arraybackslash}X}

\allowdisplaybreaks

\DeclareSymbolFontAlphabet{\mathrsfs}{rsfs}
\DeclareMathAlphabet{\mathcal}{OMS}{cmsy}{m}{n}

\usepackage{parskip}
\definecolor{darkgreen}{rgb}{0,0.5,0}

\hypersetup{
    unicode=false,          
    pdftoolbar=true,        
    pdfmenubar=true,        
    pdffitwindow=false,     
    pdfstartview={FitH},    
    pdftitle={My title},    
    pdfauthor={Author},     
    pdfsubject={Subject},   
    pdfcreator={Creator},   
    pdfproducer={Producer}, 
    pdfkeywords={keyword1} {key2} {key3}, 
    pdfnewwindow=true,      
    colorlinks=true,       
    linkcolor=red,          
    citecolor=cyan,        
    filecolor=magenta,      
    urlcolor=darkgreen,           
    linktocpage=true}

\newcommand{\ph}[1]{\phantom{#1}}

\newcommand{\be}{\begin{equation}}
\newcommand{\ee}{\end{equation}}

\newcommand{\mutp}{\widetilde{\mu}^{(2)}_{+}}
\newcommand{\mutm}{\widetilde{\mu}^{(2)}_{-}}
\newcommand{\sigmatp}{\widetilde{\sigma}^{(2)}_{+}}
\newcommand{\sigmatm}{\widetilde{\sigma}^{(2)}_{-}}
\newcommand{\di}{\mathrm{i}}

\newcommand\calO{{\mathcal{O}}}
\newcommand{\dd}{\mathrm{d}}

\newcommand{\nn}{\nonumber}

\newcommand{\tmass}{M}
\newcommand{\ADM}{\mathcal{M}}
\newcommand{\dI}{\mathrm{I}}
\newcommand{\dJ}{\mathrm{J}}
\newcommand{\dW}{\mathrm{W}}
\newcommand{\dU}{\mathrm{U}}
\newcommand{\dV}{\mathrm{V}}
\newcommand{\dM}{\mathrm{M}}
\newcommand{\dS}{\mathrm{S}}
\newcommand{\hatG}{\hat{G}}
\newcommand{\hatH}{\hat{H}}
\newcommand{\etidal}{\epsilon_\text{tidal}}

\usepackage{ulem}
\allowdisplaybreaks

\usepackage{etoolbox}
\makeatletter
\patchcmd{\frontmatter@abstract@produce}
  {\vskip200\p@\@plus1fil
   \penalty-200\relax
   \vskip-200\p@\@plus-1fil}
\makeatother

\begin{document}

\title{Tidal contributions to the full gravitational waveform to the second-and-a-half post-Newtonian order}

\author{Eve \textsc{Dones}}\email{eve.dones@obspm.fr}
\affiliation{Laboratoire Univers et Théories, Observatoire de Paris, Université PSL, Université Paris Cité, CNRS, F-92190 Meudon, France}

\author{Quentin \textsc{Henry}}\email{quentin.henry@uib.es}
\affiliation{Max Planck Institute for Gravitational Physics\\ (Albert Einstein Institute), D-14476 Potsdam, Germany}
\affiliation{Departament de Física, Universitat de les Illes Balears, IAC3 – IEEC, Crta. Valldemossa km 7.5, E-07122 Palma, Spain}

\author{Laura \textsc{Bernard}}\email{laura.bernard@obspm.fr}
\affiliation{Laboratoire Univers et Théories, Observatoire de Paris, Université PSL, Université Paris Cité, CNRS, F-92190 Meudon, France}

\date{\today}

\begin{abstract}
This paper describes the different steps to include the adiabatic tidal effects to the gravitational waveform amplitude for quasi-circular non-spinning compact binaries up to the second-and-a-half post-Newtonian (PN) order. The amplitude, that relates the two gravitational wave polarizations, is decomposed onto the basis of spin-weighted spherical harmonics of spin -2, parametrized by the two numbers $(\ell,m)$, where the modes of the waveform correspond to the coefficients of the decomposition. These modes are readily computed from the radiative multipole moments. They can be expressed in a PN-expanded form as well as in a factorized form, suitable to be directly included in effective-one-body models to describe more accurately the waveform of binary neutron stars. We also provide the energy flux and phasing evolution in time and frequency domain. The results presented in this article are collected in an ancillary file.
\end{abstract}

\pacs{04.25.Nx, 04.25.dg, 04.30.-w, 97.80.-d, 97.60.Jd, 95.30.Sf}

\maketitle

\section{Introduction}\label{sec:intro}

Gravitational waves (GW) are now routinely detected by the LIGO-Virgo-KAGRA observatories, generating a large amount of data to be processed. In the future, analysing the data will become a real challenge for the next generation of gravitational wave detectors, be they space-liked like LISA or Earth-based like the Einstein Telescope. In order to test fundamental aspects of gravity in this era of precision gravitational astronomy, even more precise gravitational wave templates are necessary. The latter are obtained by combining analytical techniques such as post-Newtonian (PN)~\cite{Blanchet:2013haa,Goldberger:2007hy,Porto:2016zng}, post-Minkowskian (PM)~\cite{Damour:2016gwp,Bern:2019crd}, black hole perturbation~\cite{Kokkotas:1999bd,Berti:2009kk} and gravitational self-force formalisms~\cite{Barack:2018yvs} with numerical relativity tools~\cite{Pretorius:2005gq,Baker:2005vv,Campanelli:2005dd}.
In this perspective, analytical modelling efforts are currently being directed in two main directions. The first one concentrates on GW modelisation in alternative theories of gravity, such as the large class of scalar-tensor theories~\cite{Bernard:2018hta,Bernard:2018ivi,Bernard:2019yfz,Bernard:2022noq,Bernard:2023eul,Trestini:2024mfs,Brax:2021qqo,Julie:2022qux}, scalar-Gauss-Bonnet~\cite{Julie:2022qux,Julie:2024fwy,Julie:2019sab,vanGemeren:2023rhh,Shiralilou:2020gah,Shiralilou:2021mfl}, Einstein-Maxwell-dilaton~\cite{Julie:2018lfp,Julie:2017rpw} or Chern-Simons~\cite{Yagi:2011xp,Yagi:2012vf} gravity theories. The second direction focuses on improving current waveform templates within the general theory of relativity. It consists for instance in adding specific effects, such as spins or tides~\cite{Faye:2006gx,Blanchet:2006gy,Flanagan:2007ix,Binnington:2009bb,Damour:2009vw}, or deriving waveforms on specific orbits, \textit{e.g.} eccentric or precessing~\cite{Damour:2004bz,Memmesheimer:2004cv,Arun:2007rg,Arun:2007sg,Arun:2009mc,Cho:2021oai,Klein:2010ti,Klein:2021jtd}, all these effects being possibly combined. Such an improvement is not only necessary for a better estimation of the parameters of the systems but it is also primordial to avoid false detections of beyond GR physics.
Over the past years, the PN community has put many efforts in including physical effects beyond the point-particle model, where the waveform now reaches 4.5PN for the phase evolution~\cite{Blanchet:2023bwj,Blanchet:2023sbv} and 3.5PN for the amplitude~\cite{Henry:2022ccf}, while the (2,2) mode is known at 4PN~\cite{Blanchet:2023bwj}. Indeed, fluxes and waveforms have been provided taking into account eccentricity~\cite{Mishra:2015bqa,Boetzel:2019nfw,Ebersold:2019kdc,Henry:2023tka}, 
spins~\cite{Blanchet:2011zv,Porto:2012as,Bohe:2015ana,Levi:2022rrq,Mishra:2016whh,Maia:2017gxn,Maia:2017yok,Siemonsen:2017yux,Cho:2021arx,Cho:2022syn,Henry:2022dzx}, 
finite-size effects~\cite{Vines:2010ca,Bini:2012gu,Damour:2012yf,Steinhoff:2016rfi,Banihashemi:2018xfb,Abdelsalhin:2018reg,HFB19,HFB20a,HFB20b,Cheung:2020sdj,Kalin:2020lmz,Mandal:2023hqa,Patil2024},
black hole horizon absorption~\cite{Poisson:1994yf,Poisson:2004cw,Goldberger:2005cd,Porto:2007qi,Saketh:2022xjb,Chiaramello:2024unv}, as well as 
electromagnetic effects~\cite{Henry:2023guc,Henry:2023len}.

In this paper, we focus on the inclusion of tidal effects in gravitational waveforms for non-spinning compact binaries. In the adiabatic approximation, tidal effects are parametrized by Love numbers characterizing the deformability of a compact object with respect to an external tidal field. While static Love numbers vanish for black holes~\cite{Damour:2009vw,Binnington:2009bb}, they are expected to be non-zero for neutron stars or exotic compact objects~\cite{Hinderer:2007mb,Cardoso:2017cfl}. Within the PN approximation, tidal effects are treated in an effective field theory (EFT) approach~\cite{Bini:2012gu} and Love numbers correspond to tidal coupling parameters. Hence, the results obtained through this approach can be applied to any compact objects (neutron stars, exotic compact objects, BHs, \textit{etc.}). 
In previous works, the conservative equations of motion (EoM) were derived at 2PN, corresponding to the next-to-next-to-leading (NNL) PN order beyond the leading (5PN) tidal contribution~\cite{HFB19}. Other diagrammatic methods, such as EFT and scattering amplitude techniques have also tackled the problem~\cite{Kalin:2020lmz,Cheung:2020sdj,Mandal:2023hqa}, and the results were shown to be equivalent in  Ref.~\cite{HFB19}. Then, the derivation of the GW flux to the 2.5PN order has been obtained, which has permitted to deduce the phase evolution to the same order, including tidal mass quadrupole, current quadrupole and mass octupole deformations~\cite{HFB20a}. Note that a gauge inconsistency between conservative~\cite{HFB19} and radiative sectors~\cite{HFB20a} led to an erroneous value of the energy flux, and consequently to the phasing. This issue has been corrected in~\cite{erratumII} and the observable quantities are in agreement with the literature, notably on the tidal mass-quadrupolar deformation terms at 2PN recently derived with the EFT method~\cite{Patil2024}. Nonetheless, the amplitude was left for future works since more PN information is required than for the derivation of the energy flux. Many of the waveform models describing NSBH or BNS systems~\cite{Nagar:2018zoe,Lackey:2016krb,Akcay:2018yyh,Lackey:2018zvw,Dietrich:2019kaq,Gamba:2023mww,Abac:2023ujg,Colleoni:2023czp} use the aforementioned PN results but the amplitude was not known to consistent PN order. The goal of this paper is to extend previous works by completing the full gravitational waveform consistently to 2.5PN order so that the waveform models can employ as much PN information as possible.

The paper is organized as follows. In Section~\ref{sec:formalism}, we introduce the general post-Newtonian multipolar post-Minkowskian (PN-MPM) formalism, we describe the form of the effective action and we outline the additional computations required compared to previous works~\cite{HFB19,HFB20a,HFB20b}. In Section~\ref{sec:intermediatecomp}, we detail the intermediate computations needed to compute the full gravitational waveform. Section~\ref{sec:results} provides the waveform amplitude modes in the ususal PN-expanded form as well as their EOB-factorized form. Some lengthy expressions are displayed in Appendices~\ref{app:moments} and~\ref{app:EOBmodes}.
We also provide our results for the GW energy flux, phase and full waveform modes, including tidal effects, as a first \emph{Mathematica} file directly available in the Supplemental Materials~\cite{SuppMaterial}. Finally, a second \emph{Mathematica} file, containing intermediate PN results for the source densities, tidal moments, potentials, source moments and the 2.5PN EoM, will be available to interested readers upon request~\cite{SuppMaterial2}.

\section{General formalism, model and conventions}\label{sec:formalism}

The PN framework allows computing waveforms, \textit{i.e.}, phase and amplitude to a given precision for general matter sources from the MPM algorithm. The PN-MPM formalism can be applied to any matter source in gravitational interaction. The model for the matter is generally used from an effective field theory (EFT) approach.

\subsection{Notations and conventions}\label{subsec:notations}

Throughout this paper, we use the signature $(-,+,+,+)$. Greek indices are four-dimensional, \textit{i.e.} $\mu = 0,1,2,3$, latin indices stand for spatial coordinates, \textit{i.e.} $i=1,2,3$ and the multi-index notation is $L=i_1\dots i_\ell$. The weighted symmetrization and antisymmetrization operators are respectively noted by parenthesis $(\dots)$ and brackets $[\dots]$ around indices. The notation $\langle\dots\rangle$ refers to the symmetric trace-free (STF) operator. The Levi-Civita tensor is noted $\epsilon_{ijk}$ with the convention $\epsilon_{123} = 1$.

We consider a binary system of compact objects of masses $m_1$ and $m_2$ within general relativity; the total mass is denoted $M = m_1 + m_2$ with $m_1 \geq m_2$. The position, velocity and acceleration of body $A = 1,2$ are written respectively $\bm{y}_A$, $\bm{v}_A$ and $\bm{a}_A$, where bold symbols represent vectors. We also define the relative separation $r_{12} \bm{n}_{12} = \bm{y}_1 - \bm{y}_2$ where $\bm{n}_{12}$ is unitary and the relative velocity $\bm{v}_{12} = \bm{v}_1 - \bm{v}_2$. The constants $G$ and $c$ are respectively the Newtonian gravitational constant and the speed of light in vacuum. Both bodies are endowed with tidal deformations parametrized by a set of mass-type and current-type tidal polarizations $\{\mu_A^{(\ell)},\sigma_A^{(\ell)}\}$. In this computation, we consider the mass-type quadrupolar and octupolar deformations as well as the current-type quadrupolar one. They are linked to the dimensionless Love numbers $k^{(\ell)}$ and $j^{(\ell)}$ through
\begin{equation}\label{eq:muAlsigmaAl}
G \mu_A^{(\ell)} = \frac{2}{(2\ell-1)!!} k_A^{(\ell)}R_A^{2\ell +1}\,, \qquad G \sigma_A^{(\ell)} = \frac{\ell-1}{4(\ell+1)(2\ell-1)!!} j_A^{(\ell)}R_A^{2\ell +1}\,,
\end{equation}
where $R_A$ is the typical radius of body $A$.
We refer to~\cite{Bini:2012gu,HFB19,HFB20a} for more details on the formalism. Within the PN approach, and using the fact that compact objects have a compactness $\mathcal{C}_A = \tfrac{G m_A}{R_A c^2}$ of the order of $\sim 1$, the tidal polarizations are at least a 5PN quantity
\begin{equation}
\mu_A^{(2)} \sim \sigma_A^{(2)} \sim \calO\left(\frac{1}{c^{10}}\right) = \calO (\etidal)\,, \qquad \mu_A^{(3)} \sim \calO\left(\frac{1}{c^{14}}\right) = \calO \left(\frac{\etidal}{c^4}\right)\,.
\end{equation}
There are two ways of thinking the PN power counting when dealing with tidal effects: the absolute and the relative ones. Throughout this paper, we use the relative way. Thus, when it is said that a quantity is required to a given PN order, it is meant that both the point-particle and tidal parts are required to that order beyond its leading order. It is illustrated by the remainder $\calO (\etidal/c^n)$.

In the center-of-mass (CoM) frame, we use the following notations: $\nu=m_1 m_2/\tmass^2$ is the symmetric mass ratio, $\delta = (m_1-m_2)/\tmass = \sqrt{1-4\nu}$ is the normalized mass difference, $\bm{x} = \bm{y}_1-\bm{y}_2$, $\bm{v} = \dd \bm{x}/\dd t$, $\bm{a} = \dd \bm{v}/\dd t$, $r = |\bm{x}| = r_{12}$ is the separation, $\bm{n} = \bm{x}/r$. We also define the following convenient combinations of the tidal polarizations
\begin{equation}\label{eq:polarpm}
\mu_\pm^{(\ell)} = \frac{1}{2}\left(\frac{m_{2}}{m_{1}}\,\mu_{1}^{(\ell)} \pm
  \frac{m_{1}}{m_{2}}\,\mu_{2}^{(\ell)}\right)\,,\qquad \sigma_\pm^{(\ell)} =
\frac{1}{2}\left(\frac{m_{2}}{m_{1}}\,\sigma_{1}^{(\ell)} \pm
  \frac{m_{1}}{m_{2}}\,\sigma_{2}^{(\ell)}\right)\,,
\end{equation}
as well as their normalized version
\begin{equation}\label{eq:musigmatilde}
\widetilde{\mu}_\pm^{(\ell)} = \left(\frac{c^2}{G \tmass}\right)^{2\ell+1}
\!\!\!G\,\mu_\pm^{(\ell)}\,,\qquad \widetilde{\sigma}_\pm^{(\ell)} =
\left(\frac{c^2}{G \tmass}\right)^{2\ell+1} \!\!\!G\,\sigma_\pm^{(\ell)}\,.
\end{equation}
Most of the computations were done using the \textit{xTensor} extension~\cite{xtensor} of the \textit{Mathematica} software. 

\subsection{Spherical harmonics decomposition of the gravitational field}\label{subsec:sphericalharmonics}

The Einstein field equations can be exactly written, by imposing the harmonic gauge condition $\partial_\nu h^{\mu\nu}=0$, as
\begin{equation}
\Box h^{\mu\nu}=\frac{16\pi G}{c^4}\tau^{\mu\nu},
\end{equation}
where $\Box$ is the flat d'Alembertian operator defined with respect to the inverse Minkowski metric $\eta^{\mu\nu}$, $h^{\mu\nu}$ is the deviation to the gothic metric $h^{\mu\nu}=\sqrt{-g}g^{\mu\nu}-\eta^{\mu\nu}$ and $g=\text{det}(g_{\mu\nu})$ is the determinant of the metric. Finally, $\tau^{\mu\nu}$ is the Landau-Lifshitz pseudotensor
\begin{equation}\label{eq:taumunu}
\tau^{\mu\nu} = \vert g\vert T^{\mu\nu}+\frac{c^4}{16\pi G}\,\Lambda^{\mu\nu}\,,
\end{equation}
where $T^{\mu\nu}$ is the stress-energy tensor and $\Lambda^{\mu\nu}$ is a function of derivatives of at least quadratic terms in the perturbed metric $h^{\mu\nu}$. Its expression in harmonic coordinates is given in Eq.~(55) of Ref.~\cite{Blanchet:2013haa}.

The transverse-traceless (TT) projection $h_{ij}^\text{TT}$ of the gravitational field can be uniquely decomposed in terms of a set of STF mass and current multipole moments $\dU_L$ and $\dV_L$, called the radiative multipole moments, as \cite{Thorne:1980ru}
\begin{align}\label{eq:hij}
h_{ij}^\text{TT} &= \frac{4G}{c^2R} \,\mathcal{P}_{ijkl} (\bm{N}) \sum^{+\infty}_{\ell=2}\frac{1}{c^\ell \ell !} \bigg[ N_{L-2} \,\dU_{klL-2}(T_R) - \frac{2\ell}{c(\ell+1)} \,N_{aL-2} \,\epsilon_{ab(k} \,\dV_{l)bL-2}(T_R)\bigg] +\mathcal{O}\left( \frac{1}{R^2} \right) \,,
\end{align}
where $R$ is the distance between the source and the observer, $\bm{N}$ is the direction of propagation of the GW and $T_R= T- R/c$ is the retarded time in some radiative gauge in which $T_R$ is asymptotically null. The quantity $\mathcal{P}_{ijkl} = \mathcal{P}_{i(k}\mathcal{P}_{l)j}-\frac{1}{2}\mathcal{P}_{ij}\mathcal{P}_{kl}$ is the TT projection operator, where $\mathcal{P}_{ij}=\delta_{ij}-N_iN_j$. The GW energy flux produced by this radiated gravitational field reads
\begin{equation}
\mathcal{F} = \sum_{\ell= 2}^{+\infty} \frac{G}{c^{2\ell+1}} \frac{(\ell+1)(\ell+2)}{(\ell-1)\ell\,\ell !(2\ell+1)!!}\bigg[\dU_L^{(1)}\dU_L^{(1)} + \frac{4\ell^2}{c^2(\ell+1)^2}\,\dV_L^{(1)}\dV_L^{(1)}\bigg]\,.
\end{equation}
On the other hand, we define the waveform polarizations as
\begin{subequations}\label{eq:hpluscross}
\begin{align}
h_+ &= \frac{1}{2}\bigl(P_i P_j - Q_i Q_j\bigr)h_{ij}^\text{TT},\\
h_\times &= \frac{1}{2}\bigl(P_i Q_j + Q_i P_j\bigr)h_{ij}^\text{TT},
\end{align}
\end{subequations}
where the vectors $(\bm{P},\bm{Q},\bm{N})$ form an orthonormal triad properly defined in \textit{e.g.} Sec. II. A. of~\cite{Henry:2022dzx}. As usual, we decompose $h_+ -\di h_\times$ in a spin-weighted spherical harmonics basis of weight -2 \cite{Kidder:2007rt},
\begin{equation}\label{eq:h}
h\equiv h_+ -\di h_\times = \sum_{l=0}^\infty \sum_{m=-\ell}^{\ell} h_{\ell m} Y^{\ell m}_{-2}(\Theta,\Phi),
\end{equation}
where the two angles $(\Theta,\Phi)$ characterize the direction of propagation $\bm{N}$. The gravitational modes are linked to the radiative moments by the relation~\cite{Faye:2012we}
\begin{equation}\label{eq:hlm}
h_{\ell m} = -\frac{2 G}{R c^{\ell+2}\ell !}\sqrt{\frac{(\ell+1)(\ell+2)}{\ell(\ell-1)}}\,\alpha_L^{\ell m} \left( U_L+\frac{2\ell}{\ell+1}\frac{\di}{c} V_L \right).
\end{equation}
Introducing a fixed orthonormal basis $(\boldsymbol{n}_0,\boldsymbol{\lambda}_0,\boldsymbol{l})$ where $\boldsymbol{l}$ is the constant vector perpendicular to the orbital plane, together with $\boldsymbol{\mathfrak{m}}_0 = (\boldsymbol{n}_0+\di \boldsymbol{\lambda}_0)/\sqrt{2}$, the projector $\alpha_L^{\ell m}$ is explicitly given by
\begin{equation}\label{eq_alphalmL_def}
\alpha_L^{\ell m} = \frac{(-)^m\,2^\frac{2+\vert m\vert}{2}\sqrt{\pi}\,\ell !}{\sqrt{(2\ell+1) (\ell+m)!\,(\ell-m)!}}\,\overline{\mathfrak{m}}_0^{\langle M}l_\text{\textcolor{white}{0}}^{L-M\rangle}\,,
\end{equation}
where the overbar denotes complex conjugation. Hence, by examining Eq.~\eqref{eq:hlm}, we see that deriving the full waveform amplitude to 2.5PN order is equivalent to requiring the knowledge of all modes $h_{\ell m}$ with $\ell \leq 7$ and $\lvert m \rvert \leq 7$. In~\Cref{table:moments}, we summarize the radiative multipole moments $\dU_L$ and $\dV_L$ that are needed, along with the orders at which they must be computed to reach the desired accuracy. 
\begin{table*}[ht]
\begin{center}
\label{tab:cycles}
\begin{tabularx}{0.95\textwidth}{cc *{5}{Y}}
\hline
  & & \multicolumn{3}{c}{Required relative PN order} \\
\hline
  $\ell$ & Moment & Gravitoelectric quadrupole $\mu^{(2)}$ & Gravitomagnetic quadrupole $\sigma^{(2)}$ & Gravitoelectric octupole $\mu^{(3)}$ \\
\hline
  2 & $\dU_{ij}$          & 2.5PN & 1.5PN & 0.5PN \\
  3 & $\dU_{ijk}$         & 2PN & 1PN & 0PN \\
  4 & $\dU_{ijkl}$        & 1.5PN & 0.5PN & - \\
  5 & $\dU_{ijklm}$       & 1PN & 0PN & - \\
  6 & $\dU_{ijklmp}$      & 0.5PN & - & - \\
  7 & $\dU_{ijklmpq}$     & 0PN & - & - \\
  \hline
  2 & $\dV_{ij}$          & 2PN & 2PN & 0PN \\
  3 & $\dV_{ijk}$         & 1.5PN & 1.5PN & - \\
  4 & $\dV_{ijkl}$        & 1PN & 1.5PN & - \\
  5 & $\dV_{ijklm}$       & 0.5PN & 0.5PN & - \\
  6 & $\dV_{ijklmp}$      & 0PN & 0PN & - \\
\hline
\end{tabularx}
\caption{Summary of the radiative moments required for the full waveform to the 2.5PN order.}\label{table:moments}
\end{center}
\end{table*} 

\subsection{From radiative to source multipole moments}\label{subsec:genmoments}

The PN-MPM formalism allows to express the radiative multipole moments, defined in Eq.~\eqref{eq:hij}, to the so-called canonical multipole moments $\{\dM_L,\dS_L\}$. 
These canonical moments are further related to the source multipole moments $\{ \dI_L,\dJ_L\}$ and the gauge ones $\{\dW_L,\text{X}_L,\text{Y}_L,\text{Z}_L \}$. The up-to-date relations between these moments can be found in \textit{e.g.} Refs.~\cite{Faye:2014fra,Blanchet:2023bwj}. In this section, we detail all the relations that are required to derive the full waveform amplitude to 2.5PN order. Note that they are valid for any type of matter distribution when only the gravitational interaction is considered. One can split the mass and current radiative moments into several pieces, as
\begin{subequations}\label{radtosource}
\begin{align}
\dU_L &= \dM_L^{(\ell)} + \dU_L^\text{tail} + \dU_L^\text{inst} + \dU_L^\text{mem}\,,\\
\dV_L &= \dS_L^{(\ell)} + \dV_L^\text{tail} + \dV_L^\text{inst}\,,
\end{align}
\end{subequations}
where the upper index $(\ell)$ refers to the $\ell^\text{th}$ time derivative. First, the tail part of the radiative moments is known for any $\ell$~\cite{Blanchet:1992br,Blanchet:1995fr} and reads
\begin{subequations}\label{eq:tails}
\begin{align}
    \dU_L^\text{tail} &= \frac{2G\ADM}{c^3} \int_0^\infty \dd \tau\left[ \ln\left( \frac{\tau}{2 b_0} \right) + \kappa_\ell \right] \dM_L^{(\ell+2)}(T_R-\tau)\,, \\
    \dV_L^\text{tail} &= \frac{2G\ADM}{c^3} \int_0^\infty \dd \tau\left[ \ln\left( \frac{\tau}{2 b_0} \right) + \pi_\ell \right] \dS_L^{(\ell+2)}(T_R-\tau)\,,
    \end{align}
\end{subequations}
where $\ADM$ refers to the Arnowitt-Deser-Misner (ADM) mass, $\{\kappa_\ell,\pi_\ell\}$ are constants of $\ell$\footnote{Their explicit expressions are given, in harmonic coordinates, by
\begin{equation*}
\kappa_\ell = \frac{2\ell^2+5\ell+4}{\ell(\ell+1)(\ell+2)}+H_{\ell-2}, \qquad \pi_\ell = \frac{\ell-1}{\ell(\ell+1)} + H_{\ell-1},
\end{equation*}
and $H_k = \sum_{j=1}^k \tfrac{1}{j}$ is the $k^\text{th}$ harmonic number.} and $b_0$ is a gauge constant inherent to the MPM algorithm associated to the coordinate shift from the near zone to the radiative zone. Next, the instantaneous pieces that contribute to the 2.5PN waveform amplitude are given by
\begin{subequations}
\begin{align}
\dU_{ij}^\text{inst} &= \frac{G}{7c^5} \biggl[ \dM_{a\langle i}^{(5)}\dM_{j\rangle a}^{\textcolor{white}{()}}-5 \dM_{a\langle i}^{(4)}\dM_{j\rangle a}^{(1)}-2 \dM_{a\langle i}^{(3)}\dM_{j\rangle a}^{(2)} +\frac{7}{3} \epsilon_{ab\langle i}\dM_{j\rangle a}^{(4)}\dS_b^{\textcolor{white}{()}}\biggr]+ \calO\left( \frac{1}{c^7} \right)\,,\\
\dU_{ijkl}^\text{inst} &= -\frac{G}{5c^3}\biggl[ 21 \dM_{\langle ij}^{(5)}\dM_{kl\rangle}^{\textcolor{white}{()}}+63 \dM_{\langle ij}^{(4)}\dM_{kl\rangle}^{(1)}+102 \dM_{\langle ij}^{(3)}\dM_{kl\rangle}^{(2)} \biggr]+ \calO\left( \frac{1}{c^5} \right)\,,\\
\dV_{ijk}^\text{inst} &= \frac{G}{10c^3}\biggl[ \epsilon_{ab\langle i}\left( \dM_{ja}^{(5)}\dM_{k\rangle b} -5 \dM_{ja}^{(4)}\dM_{k\rangle b}^{(1)}\right)-20 \,\dM^{(4)}_{\langle ij}\dS_{k\rangle}^{\textcolor{white}{(0)}} \!\!\biggr] + \calO\left( \frac{1}{c^5} \right)\,.
\end{align}
\end{subequations}
The instantaneous pieces of the other multipole moments do not contribute to the 2.5PN amplitude. Finally, the memory part (also called non-linear memory), only concerns the mass-type multipoles~\cite{Blanchet:1997jj}. In this project, we have 
\begin{subequations}\label{Umemory}
\begin{align}
\dU_{ij}^\text{mem} &= -\frac{2G}{7c^5}\int_{0}^{\infty}\dd \tau \, \dM_{a\langle i}^{(3)}(T_R-\tau)\dM_{j\rangle a}^{(3)}(T_R-\tau) +\calO\left( \frac{1}{c^7} \right)\,,\label{eq:Uijmem}\\
\dU_{ijkl}^\text{mem} &=\frac{2G}{5c^3}\int_{0}^{\infty}\dd \tau\, \dM_{\langle ij}^{(3)}(T_R-\tau)\dM_{kl\rangle}^{(3)}(T_R-\tau) +\calO\left( \frac{1}{c^5} \right)\,.
\end{align}
\end{subequations}
Now, let us link the canonical moments to the source and gauge moments. The mass quadrupole follows the relation~\cite{Blanchet:1996wx}
\begin{equation}
\dM_{ij} = \dI_{ij} + \frac{4G}{c^5}\left[ \dW^{(2)} \dI_{ij} - \dW^{(1)}\dI_{ij}^{(1)} \right] + \calO\left(\frac{1}{c^7}\right)\,.
\end{equation}
We will see in~\Cref{subsec:sourcemoments} that the gauge moment W does not contribute to the radiative quadrupole on quasi-circular orbits. The corrections for the other multipole moments are of higher PN orders. Thus, in this project, we can replace the canonical moments by the source moments, \textit{i.e.} we can substitute $\dM_L = \dI_L$ and $\dS_L = \dJ_L$. In order to derive the full waveform amplitude to 2.5PN, one needs to derive the source multipole moments to the same PN orders as for the radiative moments as displayed in~\Cref{table:moments}.

\subsection{Effective matter action}\label{subsec:effectiveaction}

The total action $S = S_\text{EH} + S_\text{m}$ is composed of the Einstein-Hilbert action for the gravitational sector endowed with a gauge fixing term,
\begin{equation}\label{eq:Sg}
S_\text{EH} = \frac{c^{3}}{16\pi G} \int \dd^{4}x \, \sqrt{-g} \left[
  R -\frac{1}{2}
  g_{\mu\nu}\Gamma^{\mu}\Gamma^{\nu} \right] \,,
\end{equation}
which ensures the harmonic coordinates condition. For the matter description, we adopt the same effective action as in Refs.~\cite{HFB19,HFB20a} to describe adiabatic tides including all the effects that can appear in the 2.5PN waveform. It is made of couplings that are quadratic in the tidal multipole moments $G_L$ (mass-type) and $H_L$ (current-type) and it reads
\begin{equation}\label{eq:Sm}
S_\text{m} = \sum_{A=1}^N \int \dd \tau_{A}\, \left( - m_A c^2 + \frac{\mu_{A}^{(2)}}{4}G^{A}_{\mu\nu}G_{A}^{\mu\nu} + \frac{\sigma_{A}^{(2)}}{6c^{2}}H^{A}_{\mu\nu} H_{A}^{\mu\nu} + \frac{\mu_{A}^{(3)}}{12} G^{A}_{\lambda\mu\nu} G_{A}^{\lambda\mu\nu} \right)\,,
\end{equation}
where $\dd\tau_A$ is the proper time of the particle $A$ along its worldline $y^{\mu}_{A}$ such that $\dd\tau_A= dt\sqrt{-g^A_{\mu\nu} v_A^{\mu}v_A^{\mu}/c^2}$. The four velocity $ u^{\mu}_{A} = \dd y^{\mu}_{A}/\dd(c\tau_{A})$ is normalized to $g^A_{\mu\nu} u_{A}^{\nu}u_{A}^{\mu} = -1$ where $g^A_{\mu\nu}$ denotes the metric evaluated at the location of body $A$, using the dimensional regularization scheme. Note that only the mass quadrupole and octupole $G_{\mu \nu}$ and $G_{\mu \nu \rho}$ and the current quadrupole $H_{\mu \nu}$ need to be considered in this work. They are defined by
\begin{subequations}\label{eq:tidalR}
\begin{align}
G^A_{\mu\nu} &= - c^2 R^A_{\mu\rho\nu\sigma} u_A^{\rho}u_A^{\sigma}\,, \\
H^A_{\mu\nu} &= 2 c^3 R^{*\, A}_{(\mu\underline{\rho}\nu)\sigma}u_A^{\rho}u_A^{\sigma}\,, \\
G^A_{\lambda\mu\nu} &= - c^2 \nabla^\perp_{(\lambda}R^A_{\mu\underline{\rho}\nu)\sigma} u_A^{\rho}u_A^{\sigma}\,.
\end{align}
\end{subequations}
where the underline notation means that the index is excluded from the symmetrization. The Riemann tensor and its dual are regularized at the location of particle $A$. We have introduced the covariant derivative projected onto the hypersurface orthogonal to the four velocity, namely $(\nabla^\perp_\mu)= (\delta_\mu^\nu + u_\mu u^\nu)\nabla_\nu$, with $\nabla^\perp_{\lambda} R^A_{\mu \nu \rho \sigma}\equiv(\nabla^\perp_{\lambda} R_{\mu \nu \rho \sigma})_A$. The tidal polarizations are constant and were introduced in Eqs.~\eqref{eq:muAlsigmaAl}.

\subsection{Post-Newtonian metric and potentials}

As for any general matter system in harmonic coordinates, the 2PN metric can be parametrized in terms of potentials as follows~\cite{Blanchet:2013haa}
\begin{subequations}\label{eq:metric}
\begin{align}
g_{00} &= -1 + \frac{2V}{c^{2}} - \frac{2V^{2}}{c^{4}}+ \frac{8}{c^{6}}
\left(\hat{X} + V_{i}V_{i} + \frac{V^{3}}{6} \right) +
\mathcal{O}\left( \frac{1}{c^{8}} \right)\,, \\ g_{0i} &=
-\frac{4V_{i}}{c^{3}} - \frac{8\hat{R}_{i}}{c^{5}} + \mathcal{O}\left(
\frac{1}{c^{7}} \right)\,, \\ g_{ij} &= \delta_{ij}\left(1 +
\frac{2V}{c^{2}} + \frac{2V^{2}}{c^{4}} \right) +
\frac{4\hat{W}_{ij}}{c^{4}} + \mathcal{O}\left( \frac{1}{c^{6}}
\right)\,,
\end{align}
\end{subequations}
where the potentials satisfy the flat spacetime wave equations
\begin{subequations}\label{eq:potentials}
\begin{align}
\Box V &= -4 \pi G \sigma\,, \\ 
\Box V_{i} &= -4  \pi G \sigma_{i}\,, \\ 
\Box \hat{W}_{ij} &= -4\pi G\bigl(\sigma_{ij} - \delta_{ij} \sigma_{kk} \bigr) -\partial_{i}V \partial_{j}V\,,\\
\Box \hat{R}_{i} &=-4 \pi G \bigl(V \sigma_{i} - V_{i} \sigma \bigr) -2 \partial_{k}V \partial_{i}V_{k} - \frac{3}{2}\partial_{t}V \partial_{i}V\,,\\
\Box \hat{X} &= -4\pi G V \sigma_{kk} + 2V_i\partial_t\partial_i V + V \partial_t^2V+\frac{3}{2}(\partial_t V)^2-2\partial_i V_j \partial_j V_i+\hat{W}_{ij}\partial_{ij}V\,.
\end{align}
\end{subequations}
The effective source densities $\sigma$, $\sigma_i$ and $\sigma_{ij}$ constitute the compact part of the potentials' sources and correspond to the different components of the stress-energy tensor as
\begin{equation}\label{eq:sourcedensities}
\sigma = \frac{T^{00}+T^{ii}}{c^2}\,,\qquad \sigma_i = \frac{T^{0i}}{c}\,, \qquad \sigma_{ij} = T^{ij}\,.
\end{equation}

\subsection{Summary of the computations}

In the rest of the paper, we detail the computations needed to complete the knowledge of the amplitude to the same 2.5PN order as the GW phase~\cite{HFB20a}. In a first stage, we compute all the intermediate quantities that are needed to then derive the full waveform. Such quantities were not needed for the previous works for two main reasons: i) the amplitude modes scale as $1/c$ of the radiative multipole moments while the GW flux scales as $1/c^2$; ii) the odd PN terms in the flux, coming from instantaneous terms (\textit{i.e.} non-tails), vanish. As a consequence, only the 2PN mass quadrupole was necessary to get the 2.5PN flux.

To obtain the radiative multipole moments at the orders given in Table~\ref{table:moments}, we need to extend previous work's computations~\cite{HFB19,HFB20a} on the following quantities:
\begin{itemize}
\item[--] the potentials $V$, $V_i$, $\hat{W}_{ij}$ and $\hat{R}_i$ in whole space, including tidal effects, and the source densities $\sigma$, $\sigma_i$ and $\sigma_{ij}$, see~\cref{subsec:potentials};
\item[--] the 2.5PN acceleration, including tidal effects, see~\cref{subsec:RR};
\item[--] the center of mass (CoM) position $G^i$, see~\cref{subsec:com};
\item[--] the tidal tensors $\hat{G}_{ij}$, $\hat{H}_{ij}$, $\hat{G}_{ijk}$, see~\cref{subsec:tidaltensors};
\item[--] the source moments to the same orders as Table~\ref{table:moments}, see~\cref{subsec:sourcemoments}.
\end{itemize}

\section{Intermediate quantities}\label{sec:intermediatecomp}

\subsection{Potentials and effective source densities}\label{subsec:potentials}

The matter stress-energy tensor, obtained by varying the effective matter action~\eqref{eq:Sm} with respect to the metric, takes the general form~\cite{Bailey:1975fe,Marsat:2014xea}
\begin{equation}\label{eq:Tmunu}
T^{\mu \nu} = \sum_{A}\left[ \mathcal{T}^{\mu \nu}_{\text{M}}\delta_A + \dfrac{1}{\sqrt{-g}}\,\partial_{\alpha}\Bigl(\mathcal{T}_{\text{D}}^{\mu \nu \alpha} \delta_{A} \Bigr) + \dfrac{1}{\sqrt{-g}}\,\partial_{\alpha \beta}\Bigl(\mathcal{T}_{\text{Q}}^{\mu \nu \alpha \beta} \delta_{A} \Bigr) + \dfrac{1}{\sqrt{-g}}\,\partial_{\alpha \beta \gamma}\Bigl(\mathcal{T}_{\text{O}}^{\mu \nu \alpha \beta \gamma} \delta_{A} \Bigr)\right]\,,
\end{equation}
where $\delta_A \equiv \delta(\bm{x} - \bm{y}_A)$ is the three dimensional Dirac distribution, the explicit values of the tensors $\mathcal{T}^{\mu \nu}_{\text{M}}$, $\mathcal{T}_{\text{D}}^{\mu \nu \alpha}$, $\mathcal{T}_{\text{Q}}^{\mu \nu \alpha \beta}$ and $\mathcal{T}_{\text{O}}^{\mu \nu \alpha \beta \gamma}$ are given in Eqs.~(2.19) of~\cite{HFB20a}. Next, following the method described in Section II. B. of~\cite{HFB20a}, we split its spatial and temporal indices in order to obtain the explicit expression of the source densities~\eqref{eq:sourcedensities}. These quantities source both the potentials and the source multipole moments. In the previous work, they were respectively required to 2PN, 1PN and 0PN orders. For the sake of the present computation, they are required respectively to 2.5PN, 2PN and 1PN orders. When expressed in terms of the potentials, they do not contain any odd PN contributions. Thus, we use the energy density $\sigma$ in Eq.~(B1a) of~\cite{HFB20a}. The explicit results for $\sigma_i$ and $\sigma_{ij}$ are not displayed in this manuscript due to their length, but can be found in the \textit{Mathematica} file available upon request~\cite{SuppMaterial2}. 

For the computations of the equations of motion at 2.5PN and of the source multipole moments, some of the potentials are required at a higher order than what was already known, namely $V$ to 2.5PN, $V_i$ to 1.5PN, $\hat{W}_{ij}$ and $\hat{R}_i$ to 0.5PN including the tidal contributions. The d'Alembertian operator in~\eqref{eq:potentials} has to be understood as the retarded propagator on the flat metric $\Box = \eta^{\mu\nu}\partial_{\mu\nu}$. Its inverse can be PN expanded and expressed in terms of Poisson integrals as
\begin{equation}\label{eq:invDalemvertian}
P(\mathbf{x},t) = \Box^{-1}_\text{ret} S(\mathbf{x},t) = -\frac{1}{4\pi} \sum_{k=0}^\infty \frac{(-)^k}{k!c^k} \frac{\dd^k}{\dd t^k}\text{PF}\int \dd^3\mathbf{x}'\vert \mathbf{x}-\mathbf{x}'\vert^{k-1} S(\mathbf{x}',t)\,,
\end{equation}
where PF is the \textit{Partie Finie} regularization scheme, to cure the divergences coming from the effective model of point masses~\cite{Blanchet:2000nu}. In particular, this computation has
to be done in the sense of distributions, for which a solution of a Laplace equation should read $\Delta(1/r_A) \propto \delta_A $, while it is not the case if the computation is made in terms of functions. The simplest example of distributional derivatives corresponds to $\partial^\text{distr}_{ab}(1/r_1) = -\tfrac{4\pi}{3}\delta_{ab}\delta_1$. The general case is given by the Gel'Fand-Shilov formula, explicitly written in \textit{e.g.} Eq.~(4.4) of~\cite{Marsat:2012fn}. It is crucial to include these contributions in the computation of the potentials and the integration of the source moments. We do not display the expressions of the ordinary part of the potentials in this paper due to their length but they will be available to the interested reader~\cite{SuppMaterial2}. We refer to \textit{e.g.}~\cite{Blanchet:1997jj} for the method to compute the ordinary part and focus hereafter on the distributional part. In particular, an interesting feature of tidal effects (as well as spin-induced multipolar deformation effects) is the fact that the potentials themselves contain a distributional part, \textit{i.e.} a Dirac distribution~$\delta_A$ appears in their expression after integration. This arises because the source densities contain double derivatives of Dirac distributions. Thus, at the lowest order, the Poisson integral of the tidal part of the source densities can lead to 
\begin{equation}
\Delta^{-1} \Bigl[ \partial_{ab} (\delta_A) \Bigr] = \partial_{ab}\left(\frac{1}{r_A}\right) = \frac{3 n_A^a n_A^b -\delta_{ab}}{r_A^3} - \frac{4\pi}{3}\delta_{ab}\,\delta_A\,.
\end{equation}
Notice that such terms are always zero when evaluated at the location of both body $A=1,2$ and after regularization. In other words, these terms do not contribute to the computation of the acceleration in~\Cref{subsec:RR} but only to the source multipole moments in~\Cref{subsec:sourcemoments}. Note also that the potentials are required to a lower PN order for the source moments than for the EoM, in particular we require $V^\text{distr}$ to 1.5PN, $V^\text{distr}_i$ to 1PN, $\hat{W}^\text{distr}_{ij}$ and $\hat{R}^\text{distr}_i$ to 0.5PN. 

We now derive the distributional terms of the potentials listed above. First, we notice that the $\calO(1/c)$ term in Eq.~\eqref{eq:invDalemvertian} does not produce distributional terms in the potentials because $\vert \mathbf{x}-\mathbf{x}'\vert^{k-1}$, which produces the distributional terms after integration by part, is absent for $k=1$.
Then, one can also check that the $\calO(1/c^3)$ terms in $V$ and $V_i$ do not produce a distributional term due to the structure of the leading order contribution to $\sigma$ and $\sigma_i$. Thus, in this section, we focus on the even terms in Eq.~\eqref{eq:invDalemvertian}. At 1PN, the compact part of potential sources, related to the source densities $\sigma$, $\sigma_i$ and $\sigma_{ij}$, takes the general form
\begin{align}\label{eq:gensource}
S(\mathbf{x},t) &= -4 \pi G f(\mathbf{x},t) \biggl[ \partial_{ab}\bigl(\delta_1 S_1^{ab}(\mathbf{x},t)\bigr) + \partial_t \partial_{a}\bigl(\delta_1 S_2^a(\mathbf{x},t)\bigr)+\partial^2_{t}\bigl(\delta_1 S_3(\mathbf{x},t)\bigr) +  \partial_{a}\bigl(\delta_1 S_4^a(\mathbf{x},t)\bigr) +  \partial_{t}\bigl(\delta_1 S_5(\mathbf{x},t)\bigr)\biggr]\nn\\
&\qquad + \delta_1 (S_6)_1 + 1\leftrightarrow 2\,.
\end{align}
The potential $P$ sourced by $S$ has a distributional part that reads up to 1PN
\begin{equation}
P^\text{distr} = -\frac{4\pi G}{3}(f)_1\left[ (S_1^{ab})_1 \left( \delta_{ab} + \frac{2v_1^a v_1^b+\delta_{ab} v_1^2}{5c^2}\right) -v_1^a (S_2^a)_1 \left( 1+\frac{3v_1^2}{5c^2}\right)+ v_1^2 (S_3)_1 \left( 1+\frac{3v_1^2}{5c^2}\right)\right]\delta_1 + 1\leftrightarrow 2\,.
\end{equation}
We see that $S_4$, $S_5$ and $S_6$ do not contribute to the distributional part of the potentials. Now, by reading the values of $S_1$, $S_2$ and $S_3$ in the values of the source densities $\sigma$, $\sigma_i$ and $\sigma_{ij}$, we deduce the distributional part of the potentials. Using general arguments regarding the Gel'Fand Shilov formula and the structure of the sources of $\hat{W}_{ij}$ and $\hat{R}_i$, one can show that the non-compact part of these potentials do not produce distributional terms. In the end, the distributional part of the potentials read
\begin{subequations}\label{eq:potdistr}
\begin{align}
V^{\text{distr}} &= \frac{2 \pi}{5}\frac{G\mu_1^{(2)}}{c^2}\hatG_1^{ab}v_1^a v_1^b \,\delta_1 + 1 \leftrightarrow 2 + \mathcal{O}\left( \dfrac{\etidal}{c^{4}} \right)\,,\\
V_i^\text{distr} &= \frac{2\pi}{5}\frac{G\mu_1^{(2)}}{c^2} v_1^i\hatG_1^{ab}v_1^a v_1^b\, \delta_1 - \frac{4\pi}{15}\frac{G\sigma_1^{(2)}}{c^2}\epsilon_{iab} v_1^a \hatH_1^{bk} v_1^k \,\delta_1+ 1 \leftrightarrow 2+ \mathcal{O}\left( \dfrac{\etidal}{c^{4}} \right)\,,\\ 
\hat{W}_{ij}^\text{distr} &= \calO\left(\frac{\etidal}{c^2}\right)\,,\\ 
\hat{R}_i^\text{distr} &= \calO\left(\frac{\etidal}{c^2}\right)\,.
\end{align}\end{subequations}
As we will see in~\Cref{subsec:sourcemoments}, these contributions are crucial for the computation of the source moments.

In order to perform some consistency checks, we write some relations between the source densities. From the conservation of the PN-expanded stress-energy tensor, namely $\nabla_\nu T^{\mu\nu} = 0$, the source densities and potentials should satisfy the following relations
\begin{subequations}\label{eq:sourcerelations}
\begin{align}
&\partial_t \sigma + \partial_i \sigma_i +\frac{1}{c^2}\bigl(\sigma \partial_t V - \partial_t \sigma_{ii} \bigr)+\frac{2}{c^4}\Bigl(2 \sigma V_i\partial_i V + \sigma_i \partial_i \hat{W} + \sigma \partial_t \hat{W} + 2\sigma_{ij}\partial_j V_i \Bigr) = \calO\left(\frac{1}{c^6}\right)\,,\\
&\partial_t \sigma_i +\partial_j \sigma_{ij} - \sigma \partial_i V +\frac{4}{c^2}\Bigl(\sigma_i\partial_t V - \sigma \partial_t V_i + \sigma_j\partial_i V_j -\sigma_j\partial_j V_i + \sigma_{ij}\partial_j V +\sigma V\partial_i V \Bigr)= \calO\left(\frac{1}{c^4}\right)\,,
\end{align}
\end{subequations}
where $\hat{W} = \delta_{ij}\hat{W}_{ij}$. When inserting the values of the source densities and the potentials, multiplying by a generic test function and integrating over the whole space, we find that the relations~\eqref{eq:sourcerelations} are respectively satisfied up to 2PN and 1PN orders. This constitutes a strong check for the expressions of the source densities in terms of the potentials.

Similarly to the relations followed by the source densities~\eqref{eq:sourcerelations}, we have also checked that the potentials follow the harmonic gauge constraints given in Eqs.~(3.7) of~\cite{HFB20a} up to 1.5PN order. Note that these relations have to be taken in the sense of distributions, which tests both the ordinary part of the potentials as well as their distributional part~\eqref{eq:potdistr}. Further checks on these distributional parts have been performed in Section~\ref{subsec:sourcemoments}.

\subsection{The 2.5PN equations of motion}\label{subsec:RR}

According to Eqs~\eqref{radtosource}, the radiative moments are obtained from the source moments, by taking several time derivatives. In particular, since the quadrupole is required to 2.5PN beyond the leading order, we need to compute the 2.5PN acceleration with tidal effects. In this section, we derive the EoM in harmonic coordinates to this order using the modified geodesic equation~\cite{Bailey:1975fe,Marsat:2014xea,HFB20a}
\begin{equation}\label{eq:modgeod}
\frac{\text{D} p_\mu}{\dd \tau} =  -\frac{1}{6} J^{\lambda \nu \rho \sigma} \nabla_\mu R_{\lambda \nu \rho \sigma} -\frac{1}{12} J^{\tau\lambda\nu\rho\sigma}\nabla_\mu\nabla_\tau R_{\lambda\nu\rho\sigma}\,.
\end{equation}
The linear momentum $p_{\mu}$, and the currents $J^{\mu \nu \rho \sigma}$ and $J^{\lambda \mu \nu \rho \sigma}$ read
\begin{align}\label{eq:pmuvalue}
p_{\mu} =& \, m c \, u_\mu + c\, \mu^{(2)} \left[ - R\indices{_\mu_\alpha_\gamma_\beta}u^{\gamma}G^{\alpha \beta} + \dfrac{3}{4c^2} u_{\mu} G^{\alpha \beta}G_{\alpha \beta} \right] +\sigma^{(2)}\left[\dfrac{4}{3}R^*_{(\mu \underline{\alpha}\gamma) \beta}  u^{\gamma}H^{\alpha \beta}  + \dfrac{1}{2c^3}H^{\alpha \beta}H_{\alpha \beta}u_{\mu}\right] \\
& + c\, \mu^{(3)} \left[ - \frac{1}{3} G^{\alpha \beta \gamma} \nabla_{\alpha} R_{\beta \mu \gamma \rho} u^{\rho} - \frac{1}{6} G_{\mu}^{\ph\mu \beta \gamma}  u^{\kappa}\nabla_{\kappa} R_{\beta \rho \gamma \sigma} u^{\rho}u^{\sigma} + \frac{1}{4c^2} G_{\alpha \beta \gamma}G^{\alpha \beta \gamma} u_{\mu}\right] \,,\nn\\
J^{\mu \nu \rho \sigma} =& \, - 3 c^2 \mu^{(2)} u^{[ \mu}G^{\nu ] [ \rho} u^{\sigma ]} + c \,\sigma^{(2)}\left( \varepsilon\indices{^\mu^\nu_\alpha_\beta}u^{\alpha}H^{\beta [\rho}u^{\sigma ]} + \varepsilon\indices{^\rho^\sigma_\alpha_\beta}u^{\alpha}H^{\beta [\mu}u^{\nu ]} \right)\,,\\
J^{\lambda \mu \nu \rho \sigma} =& \, -2 c^2 \mu^{(3)}  
u^{[ \mu}G^{\nu ] \lambda [ \rho}u^{\sigma ]}\,,
\end{align}
where we recall $\nabla_{\mu}^{\perp}=\perp_{\mu}^{\nu}\nabla_{\nu} $ and $\perp_{\mu}^{\nu} =\delta_{\mu}^{\nu} + u_{\mu} u^{\nu}  $. We first perform a (3+1) splitting as well as a PN expansion of Eq.~\eqref{eq:modgeod} which can be recasted into the following form
\begin{equation}\label{dPdtF}
\frac{\dd P^i}{\dd t} = F^i\,,
\end{equation}
where the quantities $P^i = P^i_\text{M} + P^i_\text{Q} + P^i_\text{O}$ and $F^i = F^i_\text{M} + F^i_\text{Q} + F^i_\text{O}$ respectively contain monopolar, quadrupolar and octupolar contributions. First, we recover known results from Refs.~\cite{Blanchet:1998vx,Marsat:2012fn} for the monopolar part up to 2PN
\begin{subequations}\label{PM}
\begin{align}
P^i_\text{M} = m \Biggl\{& v^i +\frac{1}{c^2}\biggl[ \frac{1}{2}v^2 v^i + 3V v^i -4 V_i \biggr] \nn \\
&  \quad + \frac{1}{c^4} \biggl[ \frac{3}{8}v^4 v^i +\frac{7}{2}V v^2 v^i - 4 V_j v^i v^j -2V_i v^2 +\frac{9}{2} V^2 v^i -4 V V_i +4 \hat{W}_{ij}v^j -8\hat{R}_i\biggr] \Biggr\} + \calO\left( \frac{1}{c^6}\right) \,,\\
F^i_\text{M} = m \Biggl\{&  \partial_i V +\frac{1}{c^2}\biggl[-V \partial_i V +\frac{3}{2} \partial_i V v^2 -4\partial_i V_j v^j \biggr] + \frac{1}{c^4} \biggl[\frac{7}{8}\partial_i V v^4 -2\partial_i V_j v^j v^2+\frac{9}{2}V\partial_i V v^2  \nn \\
& \quad +2\partial_i\hat{W}_{jk}v^j v^k -4V_j\partial_i V v^j-4V \partial_i V_j v^j- 8\partial_i \hat{R}_j v^j +\frac{1}{2}V^2\partial_i V +8 V_j\partial_iV_j +4 \partial_i \hat{X}\biggr]\Biggr\} + \calO\left( \frac{1}{c^6}\right)\,.
\end{align}
\end{subequations}
Then the quadrupolar contributions read
\begin{subequations}\label{PQ}
\begin{align}
P^i_\text{Q} =&\, \frac{1}{c^2}\biggl[ \mu^{(2)}\biggl(\frac{3}{4}\hatG_{ab}\hatG_{ab} v^i + \hatG_{ab}\partial_{ab}V v^i -\hatG_{ia}\partial_t\partial_a V - \hatG_{ib} v^a \partial_{ab}V -2\hatG_{ab}\partial_{ab} V_i  -2 \hatG_{ab}v^a\partial_{ib}V +2\hatG_{ab}\partial_{ib}V_a \biggr) \nn \\
& \qquad  + \frac{4}{3} \sigma^{(2)} \epsilon_{ibk}\hatH_{ab} \partial_{ka}V \biggr] + \calO\left(\frac{\etidal}{c^4}\right)\,,\\
F^i_\text{Q} =&\, \frac{\mu^{(2)}}{2}\hatG_{ab}\partial_{iab}V + \frac{1}{c^2}\Biggl\{ \mu^{(2)}\biggl[ \hatG_{ib}\partial_a V \partial_{ab}V + \hatG_{ab}\left(\frac{3}{4} \hatG_{ab} \partial_i V - v^a\partial_t\partial_{ib}V + 2 \partial_t\partial_{ib}V_a  -2 \partial_{ab}V \partial_i V  \right. \nn \\
& \left. -4\partial_a V\partial_{ib}V +\frac{3}{4}v^2\partial_{iab}V -\frac{3}{2} V \partial_{iab}V+ 2v^j \partial_{ija}V_b -\frac{3}{2}v^a v^j\partial_{ijb}V -2 v^j\partial_{iab}V_j\right)\biggr] \nn \\
& + \frac{4}{3}\sigma^{(2)}\epsilon_{abk}\biggl(\hatH_{aj}\partial_{ijk}V_b + \hatH_{bj} v^a\partial_{ijk}V \biggr) \Biggr\} + \calO\left(\frac{\etidal}{c^4}\right)\,.
\end{align}
\end{subequations}
We omit here the 2PN terms for sake of length but we make them available to the interested reader~\cite{SuppMaterial2}. Finally, the leading order mass-octupolar part is given by
\begin{subequations}\label{PO}
\begin{align}
P^i_\text{O} =& \, \calO\left(\frac{\etidal}{c^6}\right)\,, \\
F^i_\text{O} =&\, \frac{\mu^{(3)}}{6}\hatG_{abk}\partial_{iabk}V + \calO\left(\frac{\etidal}{c^6}\right)\,.
\end{align}
\end{subequations}
To compute the equations of motion for body $A$, one has to evaluate each quantity appearing in Eqs.~\eqref{dPdtF}--\eqref{PO} at the location of body $A$, namely we replace the mass $m$ by $m_A$, the velocity $v^i$ by $v^i_A$, $\mu^{(2)}$ by  $\mu^{(2)}_A$, \textit{etc.}, and evaluate the tidal tensors, the potentials and their derivatives at the location of body $A$. 
From these expressions and as anticipated in~\Cref{subsec:potentials}, we see that we need $V$ to 2.5PN, $V_i$ to 1.5PN, $\hat{W}_{ij}$ and $\hat{R}_i$ to 0.5PN orders including their tidal contributions. We also have the occurrence of the potential $\hat{X}$ only through its spatial derivative, $(\partial_i\hat{X})$, needed to 0.5PN order. In practice, it is sufficient and easier to compute it directly regularized at point $A$, namely $(\partial_i\hat{X})_A$, using Eq.~(5.17a) of~\cite{Blanchet:2000nu}. The derivation of the other potentials is explained in~\Cref{subsec:potentials}.

After injecting the potentials, we get the full EoM up to 2.5PN in harmonic coordinates. The full EoM to 2.5PN are made available in a \textit{Mathematica} file for the interested reader~\cite{SuppMaterial2}. Note however that the EoM differ at the 2PN order from the one derived in Ref.~\cite{HFB19}. Indeed in the latter paper, the EoM were derived using a Fokker Lagrangian approach but in a different gauge than the harmonic one used in the present work. More precisely, one of the PN potentials entering in the Fokker Lagrangian, namely $\hat{X}$, had been replaced by its value on shell, which is equivalent to applying a gauge transformation to the particle positions~\cite{Damour:1990jh}.
As a consequence, the correction to add to the relative acceleration given in Eq.~(C2) of~\cite{HFB19} to get the correct result in harmonic gauge is given by
\begin{equation}\label{acccorr}
\delta a^i = \frac{6 G^3 \tmass^2}{c^4 r^8}\Bigl(\mu_{+}^{(2)}+\delta\,\mu_{-}^{(2)}\Bigr) \left[\left( 4\frac{G \tmass}{r} +63(nv)^2 -7v^2 \right)n^i-14 (nv)v^i\right]\,.
\end{equation}
Note that this also modifies the conserved quantities published in~\cite{HFB19} and in particular the conserved energy and angular momentum in the CoM frame. However, their expressions in the case of circular orbits remain unchanged since observables quantities are gauge independent. In~\Cref{sec:correctedConservative}, we give the expressions of the CoM Lagrangian and associated conserved quantities in the harmonic gauge consistent with the one considered here.

As a consistency check, the radiation reaction term of the relative acceleration in the CoM frame and on quasi-circular orbits can also be computed from the energy flux balance equation
\begin{equation}\label{eq:flux_balance_eq}
\frac{\dd E}{\dd t} = -\mathcal{F}\,,
\end{equation}
using the conserved energy~\eqref{eq:Etidal} and the GW energy flux $\mathcal{F}$~\cite{HFB20a}. It reads
\begin{equation}
\frac{\dd v^i}{\dd t} = -r\omega^2 n^i- \frac{32 G^3 \tmass^3\nu}{5 c^5 r^4}\left[ 1+\frac{6G}{\nu r^5}\left( (1+6\nu) \mu_+^{(2)} +\delta \mu_-^{(2)}  \right) \right]v^i + \calO\left(\frac{1}{c^6},\frac{\etidal}{c^6}\right)\,,
\end{equation}
where $\omega$ is the orbital frequency, whose value on circular orbits is given in Eq.~\eqref{eq:omega2}. In the quasi-circular limit, both methods to compute the radiation reaction term in the EoM yield the same result.

\subsection{Center of mass position}\label{subsec:com}

In Ref.~\cite{HFB19}, the CoM position was derived only up to 1PN order. Indeed, the 2PN tidal correction was not required for the computation of the 2PN GW energy flux, due to the structure of the leading order source mass quadrupole. In the present work, since we need the source mass octupole and current quadrupole to 2PN, the knowledge of the full 2PN CoM position is necessary. It is obtained by imposing $\dd^2 G^i/\dd t^2 = 0$. In this section, we directly provide the result for the CoM position, and we refer the reader to Ref.~\cite{deAndrade:2000gf} for an alternative Lagrangian method to compute the same quantity.
We split the CoM position in its point-particle and tidal contributions as $G^i = G^i_\text{pp} + G^i_\text{tidal}$. The point-particle part is provided in \textit{e.g.}, Eq. (4.5) of~\cite{deAndrade:2000gf}, and the tidal one reads
\begin{align}\label{eq:Gitidal}
G^i_\text{tidal} &= \frac{G^2 m_2^2}{r_{12}^5}\Biggl\{ \frac{3\mu_{1}^{(2)}}{2 c^2}
 \left(3 n_{12}^{i}- \frac{y_{1}^{i}}{r_{12}} \right) \nn \\
& +\frac{1}{c^4}\Biggl[\mu_{1}^{(2)}\Biggl[ n_{12}^i\Big( -69 (n_{12}v_{12})^2 + 15 v_{12}^2 + 18(n_{12}v_{12})(n_{12}v_{1}) - 9(n_{12}v_{1})^2+\frac{9}{4}v_1^2\Big) \nn \\
& \quad + \frac{3y_1^i}{4r_{12}}\left( 24(n_{12}v_{12})^2 -24 (n_{12}v_{12})(n_{12}v_{1}) -6(n_{12}v_{1})^2+6(v_{12}v_{1})-v_{1}^2\right) + \frac{3}{2}v_{12}^i\Big( 11(n_{12}v_{12})-3(n_{12}v_{1})\Big)  \nn \\
& \quad + \frac{G}{r_{12}} \left( m_1  \left( -\frac{285}{8} n_{12}^i +3\frac{y_1^i}{r_{12}} \right)+ m_2\left( -\frac{181}{8} n_{12}^i +\frac{21y_1^i}{2r_{12}} \right) \right)\Biggr] \nn \\
& \quad + \sigma_1^{(2)} \left[ 8\Bigl( (n_{12}v_{12})^2 - v_{12}^2 + 3(v_{12}v_1) \Bigr) n_{12}^i -24 (n_{12}v_1) v_{12}^i + 12 \Bigl( v_{12}^2 - (n_{12}v_{12})^2 \Bigr)\frac{y_1^i}{r_{12}}\right]\Biggr]\Biggr\}\nn \\ 
& + 1\leftrightarrow2 + \mathcal{O}\left(\frac{\epsilon_\text{tidal}}{c^{6}} \right)\,.
\end{align}
By solving iteratively $G^i=0$, we can express the positions $y_A^i=\left(y_A^i\right)_\text{pp}+\left(y_A^i\right)_\text{tidal}$ and the velocities $v_A^i=\left(v_A^i\right)_\text{pp}+\left(v_A^i\right)_\text{tidal}$ of the two bodies in the CoM frame. Their tidal part to 2PN order read
\begin{subequations}\label{eq:yiviCoM}
\begin{align}
\left(y_1^i\right)_\text{tidal} =& - \frac{3 G^2 M \nu}{2 c^2 r^5} \Biggl\{\Bigl(\delta\,\mu_{+}^{(2)} + 5 \mu_{-}^{(2)} \Bigr) n^{i} + \frac{1}{c^2}\biggl[ \biggl(\frac{G M}{3 r} \delta\,\mu_{+}^{(2)} (-47 + 6 \nu)  - \frac{G M}{6 r} (191 + 30 \nu) \mu_{-}^{(2)}  \nn \\
&+ \left(- 3 \delta\,\mu_{+}^{(2)} (1 + 3 \nu) + \mu_{-}^{(2)} (-89 + 75 \nu) + 8 \delta\, \sigma_{+}^{(2)} + \frac{8}{3} \sigma_{-}^{(2)}   \right) (n v)^2 \nn\\
& + \left(- \delta\, \mu_{+}^{(2)}(4 + 3 \nu) - 3 \mu_{-}^{(2)} (-8 + 5 \nu) - 24 \delta\, \sigma_{+}^{(2)} + \frac{40}{3} \sigma_{-}^{(2)}  \right) v^2\biggr)n^{i} \nn \\
& + \biggl(3 \delta \,\mu_{+}^{(2)}+ 19 \mu_{-}^{(2)} + 16 \delta \,\sigma_{+}^{(2)} -16 \sigma_{-}^{(2)} \biggr) (n v) v^{i}\biggr]\Biggr\} + \calO\left( \frac{\epsilon_\text{tidal}}{c^6}\right)\,,\\
\left(v_1^i\right)_\text{tidal} =& - \frac{3 G^2 M \nu}{2 c^2 r^6} \Biggl\{\Bigl(6 \delta\, \mu_{+}^{(2)} -30 \mu_{-}^{(2)} \Bigr)  (n v) n^{i} + \Bigl(\delta\, \mu_{+}^{(2)} + 5 \mu_{-}^{(2)} \Bigr) v^{i} \nn \\
& + \frac{1}{c^2}\biggl[ \biggl( \left(24 \delta\, \mu_{+}^{(2)} (-2 + \nu) - 8 \mu_{-}^{(2)} (-74 + 15 \nu)  + 64 \delta \,\sigma_{+}^{(2)}- \frac{64}{3} \sigma_{-}^{(2)}   \right) (n v)^2 \nn \\
&+ \left( - 3 \delta\,\mu_{+}^{(2)} (-7 + 4 \nu) + (-277 + 60 \nu) \mu_{-}^{(2)} + 32 \delta\, \sigma_{+}^{(2)} - \frac{224}{3} \sigma_{-}^{(2)}    \right) v^2 \nn \\
& +\frac{G\tmass}{r} \left( \frac{1}{3} \delta\,\mu_{+}^{(2)} (95 + 78 \nu) -  \frac{1}{6} \mu_{-}^{(2)}(-2063 + 750 \nu) + 16 \delta\, \sigma_{+}^{(2)} - 16 \sigma_{-}^{(2)} \right)  \biggr) (nv) n^{i} \nn \\
& + \biggl( \left(- 3 \delta\,\mu_{+}^{(2)} (7 + 3 \nu) + \mu_{-}^{(2)} (-203 + 75 \nu) - 88 \delta \,\sigma_{+}^{(2)} + \frac{296}{3} \sigma_{-}^{(2)}   \right) (n v)^2  \nn\\
&+ \left(-  \delta \,\mu_{+}^{(2)} (1 + 3 \nu) + (43 - 15 \nu) \mu_{-}^{(2)}- 8 \delta\, \sigma_{+}^{(2)} - \frac{8}{3} \sigma_{-}^{(2)} \right)v^2 \nn \\
& + \frac{G M}{r} \left(\frac{1}{3} \delta\,\mu_{+}^{(2)}  (-119 + 6 \nu) - \frac{5}{6} (61 + 6 \nu) \mu_{-}^{(2)} - 16 \delta\,\sigma_{+}^{(2)} + 16 \sigma_{-}^{(2)} \right) \biggr) v^{i}\biggr]\Biggr\} + \calO\left( \frac{\epsilon_\text{tidal}}{c^6}\right)\,,
\end{align}
\end{subequations}
while $\left(y_2^i\right)_\text{tidal}$ and $\left(v_2^i\right)_\text{tidal}$ are given by $1\leftrightarrow2$. We recall that the notations are explained in Section~\ref{subsec:notations}. The point-particle contributions $\left(y_A^i\right)_\text{pp}$ and $\left(v_A^i\right)_\text{pp}$ are known up to 4PN, see Eqs.~(B.4) in~\cite{Bernard:2017ktp}.

\subsection{Tidal tensors}\label{subsec:tidaltensors}

As explained in Section~II.~B. of~\cite{HFB20a}, it is convenient to use the projection of the tidal tensors on the worldline tetrad, see Eqs~(2.22) of~\cite{HFB20a}. In the present work, the tidal tensors projected on the tetrad, $\hatG_{ij}$, $\hatH_{ij}$ and $\hatG_{ijk}$, are required to respectively 2.5PN, 2PN and 0.5PN orders in the point-particle approximation.

Similarly to Sec.~\ref{subsec:RR}, we perform a (3+1) decomposition coupled to a PN expansion of Eqs.~\eqref{eq:tidalR}, to get an expression of the tidal moments in terms of the potentials, evaluated at the location of body $A$. As mentioned in~\Cref{subsec:effectiveaction}, we use dimensional regularization to treat the self-field divergences that appear. In the case of point-particles, \textit{i.e.} without any additional finite-size effects such as tides, spins, \textit{etc.}, Hadamard regularization is equivalent to dimensional regularization up to 2.5PN order. Furthermore, using general arguments on the structure of the integrals appearing in the calculations, it was shown in~\cite{Bini:2012gu,HFB20a} that up to now tidal effects could be treated using Hadamard regularization only. It is \textit{a priori} no more the case in the present work as we need to compute the current-type tidal tensor $\hat{H}_{ij}$ to 2PN order. Indeed, the expression of this tensor involves four complicated potentials: $(\partial_{ij}\hat{R}_k)_1$ to 1PN, $(\partial_{ij}\hat{X})_1$, $(\partial_{ij}\hat{Y}_k)_1$ and $(\partial_{ij}\hat{Z}_{ab})_1$ at Newtonian order. Sections IV. A. and IV. B. of Ref~\cite{Blanchet:2003gy} detail how to deal with such self-field regularization. For practical purpose, it is sufficient to compute first the full Hadamard-regularized value and then to add the difference between dimensional and Hadamard regularization, in order to obtain the fully $d$-dimensional regularized value. For the potentials we are dealing with, the problem has already been dealt with in Ref.~\cite{Marsat:2012fn}. In particular, among the four problematic potentials, only $(\partial_{ab}\hat{Y}_{i})_1$ gives a difference between both regularizations, and in particular induces a pole of the form
\begin{equation}\label{eq:Ypole}
\mathcal{D}(\partial_{ij}\hat{Y}_k)_1 \propto \frac{1}{\varepsilon} \, \left(\alpha_1 v_1^a -\alpha_2 v_2^a\right)
\partial^1_{ijka}\left( \frac{1}{r_{12}}\right) + \calO(\varepsilon^0)\,,
\end{equation}
where $\varepsilon = d-3$, $\alpha_{1,2}$ are some numerical coefficients\footnote{Note that we found a different result from~\cite{Marsat:2012fn} for the pole so we have not put the explicit expression here. However, as the two results share the same symmetry feature, it does not spoil the fact that it will not contribute to the tidal tensors.} and $\partial^1_i$ is the partial derivative with respect to $y_1^i$. In~\cite{Marsat:2012fn}, the authors show that it does not contribute to the spin EoM because it is a symmetric tensor contracted with the antisymmetric spin tensor. Let us show that we are in the same situation in the present computation. The tidal tensor $H_{\mu\nu}$, as expressed in Eq.~\eqref{eq:tidalR}, involves the Levi-Civita tensor which cannot be properly defined in arbitrary dimensions. Fortunately, the problem of defining current-type moments in $d \in \mathbb{C}$ dimensions has been tackled in~\cite{Henry:2021cek}. In particular, the moment $H_{k\vert ji}$, which corresponds to the Hodge dual of $H_{ij}$ in the limit $d\rightarrow 3$, is introduced. Amongst the many symmetries of such a moment, $H_{k\vert ji}$ is antisymmetric over $k$ and $j$. More precisely, the potential $\hat{Y}_i$ enters in $H_{k\vert ji}$ through the contribution $H_{k\vert ji}^{\hat{Y}_i} \propto \partial_{i[j}\hat{Y}_{k]}$. However, both the pole and the finite part of $\mathcal{D}(\partial_{ij}\hat{Y}_k)_1$ are symmetric over $j$ and $k$. Thus, when replacing Eq.~\eqref{eq:Ypole} in $H_{k\vert ji}$, one finds that the difference between dimensional and Hadamard regularizations yields zero. In other words, one can safely use Hadamard regularization for the computation of all the tidal tensors.

Using the definitions~\eqref{eq:tidalR}, we first decompose the tidal tensors in their spatial and temporal indices, project them onto the tetrad given in Section II.~B. of~\cite{HFB20a} and express the tidal tensors in terms of the PN potentials. Then, we replace the potentials by their value regularized at point 1. They read~\footnote{The cross product is denoted, \textit{e.g.} $(n_{12} \times v_{12})_i$, and the mixed product, \textit{e.g.} $(v_2,n_{12},v_1)= v_2 . (n_{12} \times v_{1})$}
\begin{subequations}\label{tidaltensor}
\begin{align}
\hat{G}_1^{ij} =& \frac{G m_{2}{}}{r_{12}{}^3} \Biggl\{3 n_{12}{}_{\langle i} n_{12}{}_{j \rangle}\nn \\
& + \frac{1}{c^2}\biggl[n_{12}{}_{\langle i} n_{12}{}_{j \rangle} \biggl( 6 v_{1}^2 -  \tfrac{15}{2} (n_{12} v_{2})^2 - 12 (v_{1} v_{2}) + 6 v_{2}^2 - \frac{15 G m_{1}{}}{2 r_{12}{}} -  \frac{9 G m_{2}{}}{r_{12}{}}\biggr) - 9 n_{12}{}_{\langle i}  v_{1}{}_{j \rangle}(n_{12} v_{1}) \\
& + 6 n_{12}{}_{\langle i} v_{1}{}_{j \rangle} (n_{12} v_{2}) + 3 v_{1}{}_{\langle i} v_{1}{}_{j \rangle} + 12 n_{12}{}_{\langle i}  v_{2}{}_{j \rangle}(n_{12} v_{1}) - 6 n_{12}{}_{\langle i} v_{2}{}_{j \rangle} (n_{12} v_{2}) - 6 v_{1}{}_{\langle i} v_{2}{}_{j \rangle} + 3 v_{2}{}_{\langle i} v_{2}{}_{j \rangle}\biggr]\Biggr\} + \calO\left( \frac{1}{c^4}\right) \nn \,,\\
\hat{H}_1^{ij} = & \frac{G m_{2}{}}{r_{12}{}^3} \Biggl\{12 (n_{12} \times v_{12})_{\langle  i} n_{12}{}_{j \rangle} \nn \\
& + \frac{1}{c^2}\biggl[(n_{12} \times v_{12})_{\langle i} n_{12}{}_{j \rangle} \biggl( 12 v_{1}^2 - 30 (n_{12} v_{2})^2 - 12 (v_{1} v_{2}) + 12 v_{2}^2 - \frac{4 G m_{1}{}}{r_{12}{}} -  \frac{12 G m_{2}{}}{r_{12}{}}\biggr) \nn \\
& - 6 (n_{12} \times v_{12})_{\langle i}  v_{1}{}_{j \rangle}(n_{12} v_{1}) + 12 (n_{12} \times v_{12})_{\langle i}  v_{2}{}_{j \rangle}(n_{12} v_{1}) - 6 n_{12}{}_{\langle i} v_{1}{}_{j \rangle} (v_2,n_{12},v_1)\biggr]\Biggr\}+ \calO\left( \frac{1}{c^4}\right)\,,\\
\hat{G}_1^{ijk} =& - \frac{15 G m_{2} n_{12\langle i} n_{12j } n_{12k\rangle}}{r_{12}^4}+ \calO\left( \frac{1}{c^2}\right)\,.
\end{align}
\end{subequations}
Once again, we only display the 1PN mass and current tidal quadrupoles and the 0.5PN mass octupole. The full results are displayed in the second \emph{Mathematica} file available upon request~\cite{SuppMaterial2}. Be careful that Eqs.~\eqref{tidaltensor} differ from the explicit expressions for the tidal tensors computed \textit{off-shell} and not projected on the tetrad that were given in~\cite{HFB19}.

\subsection{Source moments}\label{subsec:sourcemoments}

The source multipole moments are obtained by a matching between the multipole expansions respectively performed in the near and exterior zones~\cite{Blanchet:1998in,Poujade:2001ie}. They are given as integrals over the stress-energy (or Landau-Lifshitz) pseudotensor~\eqref{eq:taumunu}, and read
\begin{subequations}\label{eq:genMultMom}
\begin{align}
\dI_{L} &= \underset{B=0}{\mathrm{FP}} \int \dd^{3}\mathbf{x} \,\vert \widetilde{\mathbf{x}}\vert^B \left[ \hat{x}_{L} \overline{\Sigma}_{[\ell]} - \dfrac{4(2\ell+1)}{c^2(\ell+1)(2\ell+3)} \hat{x}_{iL} \overline{\Sigma}^{(1)}_{i[\ell+1]} + \dfrac{2(2\ell+1)}{c^{4}(\ell+1)(\ell+2)(2\ell+5)}  \hat{x}_{ijL} \overline{\Sigma}^{(2)}_{ij[\ell+2]} \right]\,,\label{eq:IL}\\
\dJ_{L} &= \epsilon_{ab \langle i_{\ell}} \underset{B=0}{\mathrm{FP}} \int \dd^{3}\mathbf{x}\,\vert \widetilde{\mathbf{x}}\vert^B \left[ \hat{x}_{L-1 \rangle a} \overline{\Sigma}_{b[\ell]} - \dfrac{2\ell+1}{c^2(\ell+2)(2\ell+3)} \hat{x}_{L-1 \rangle ac} \overline{\Sigma}^{(1)}_{bc[\ell]} \right]\,,\label{eq:JL}
\end{align}
\end{subequations}
where $\widetilde{\mathbf{x}} = r/r_0$ with $r_0$ corresponding to an infrared cutoff, and
\begin{equation}
\overline{\Sigma}_{A[\ell]} = \sum_{k=0}^{\infty} \alpha_\ell^k\left( \frac{r}{c}\frac{\partial}{\partial t}\right)^{2k} \overline{\Sigma}_A\,,
\end{equation}
with $A=\{\varnothing,i,ij\}$, $\alpha_\ell^k = \tfrac{1}{2^k k!}\frac{(2\ell+1)!!}{(2\ell+2k+1)!!}$, the overbar notation refers to the PN expansion and the $\Sigma$'s are defined as the $3+1$ decomposition of the stress-energy pseudo-tensor $\tau^{\mu\nu}$,
\begin{align}\label{eq:Sigma}
\Sigma \equiv \frac{\tau^{00}+\tau^{ii}}{c^2}\,,\qquad 
\Sigma_i \equiv \frac{\tau^{0i}}{c}\,,\qquad
\Sigma_{ij} \equiv \tau^{ij}\,.
\end{align}
Finally, in Eqs.~\eqref{eq:genMultMom}, the finite part ($\mathrm{FP}$) operator is present to cure the infra-red (IR) divergences of the PN-expanded integrand in the source multipole moments at spatial infinity. More details on how to deal in practice with the IR regularization can be found in Ref.~\cite{Marchand:2020fpt}. 

The general form of the mass-type source moments as integrals over the potentials and source densities up to 2PN order is given in Eqs.~(4.4) of~\cite{HFB20a}. We provide here the expression for the current-type source moments to 2PN in 3 dimensions. This expression has not been published before, although it was used in~\cite{Blanchet:2008je} to derive the current quadrupole to 2.5PN. It takes the following form,
\begin{equation}
\dJ_L = \mathrm{VI}_L + \mathrm{VII}_L + \mathrm{VIII}_L + \mathrm{TI}_L +
\mathrm{TII}_L \,,
\end{equation}
where each block reads
\begin{subequations}\label{eq:JLblocks}
\begin{align}
\mathrm{VI}_L &= \epsilon_{ab\langle i_\ell}
\mathop{\mathrm{FP}}_{B=0} \int \dd^3\mathbf{x}\,\vert \widetilde{\mathbf{x}}\vert^B \hat{x}_{L-1\rangle a}  \biggl\{ \sigma_b + \frac{1}{c^2} \Big[2( \sigma_b V -  \sigma V_b ) + \frac{1}{\pi G} \Big(\partial_i V \partial_b V_i + \frac{3}{4} \partial_t V \partial_b V - \frac{1}{2} \Delta (V V_b) \Big) \Big] \nonumber \\ 
& \quad + \frac{1}{c^4} \Big[2( \sigma_b V^2 - 2 \sigma \hat{R}_b +  V_b \sigma_{ii} +  \hat{W}_{bi} \sigma_i +  V_i \sigma_{bi}) + \frac{1}{\pi G}\Big( \frac{1}{2} V_b \partial_t^2 V  - \frac{1}{2} V \partial_t^2 V_b + \partial_t V \partial_t V_b  - 2 V_i \partial_i \partial_t V_b
\nonumber \\ & \qquad  \quad + \frac{3}{2} V \partial_t V \partial_b V - 
V_b \partial_i V \partial_i V+ \frac{3}{2} V_i \partial_b V \partial_i V + 
2 \partial_i V \partial_b \hat{R}_i - \hat{W}_{ij}\partial_{ij} V_b + \partial_t \hat{W}_{bi} \partial_i V - \partial_i V_j \partial_b \hat{W}_{ij} \nonumber \\ 
& \qquad \quad +\partial_i \hat{W}_{bj} \partial_j V_i -\frac{1}{2} \Delta (V^2 V_b) - \frac{1}{2} \Delta (\hat{W} V_b) + \frac{1}{2} \Delta (\hat{W}_{bi} V_i) - \Delta (V \hat{R}_b)\Big) \Big]\biggr\}\,, \\
\mathrm{VII}_L &= \frac{1}{2c^2(2\ell +3)} \epsilon_{ab\langle i_\ell} \mathop{\mathrm{FP}}_{B=0}\, \frac{\dd^2}{\dd t^2}\int \dd^3\mathbf{x}\,\vert \widetilde{\mathbf{x}}\vert^B \biggl\{ \vert \mathbf{x}\vert^2 \hat{x}_{L-1\rangle a} \biggl[ \sigma_b + \frac{1}{c^2}\Big[2( \sigma_b V -  \sigma V_b ) + \frac{1}{\pi G} \Big( \partial_i V \partial_b V_i + \frac{3}{4} \partial_t V \partial_b V \Big)\Big] \biggr] \nonumber \\ 
& \quad - \frac{1}{\pi G c^2} \biggl[(2\ell+3) \hat{x}_{L-1\rangle a}( V V_b ) + \frac{1}{2} \partial_i \Bigl[\partial_i (V V_b) \vert \mathbf{x}\vert^2 \hat{x}_{L-1\rangle a} - V V_b \partial_i (\vert \mathbf{x}\vert^2 \hat{x}_{L-1\rangle a})\Bigr] \biggr] \biggr\}\,,\\ 
\mathrm{VIII}_L &= \frac{1}{8c^4(2\ell +3)(2\ell +5)}  \epsilon_{ab\langle i_\ell} \mathop{\mathrm{FP}}_{B=0}\, \frac{\dd^4}{\dd t^4}\int \dd^3\mathbf{x}\,\vert \widetilde{\mathbf{x}}\vert^B \hat{x}_{L-1\rangle a} \vert \mathbf{x}\vert^4 \sigma_b \,,\\
\mathrm{TI}_L &= - \frac{(2\ell +1)}{c^2(\ell +2)(2\ell +3)} \epsilon_{ab\langle i_\ell}
\mathop{\mathrm{FP}}_{B=0}\, \frac{\dd}{\dd t}\int \dd^3\mathbf{x}\,\vert \widetilde{\mathbf{x}}\vert^B\hat{x}_{L-1\rangle ak}\biggl\{ \sigma_{bk} + \frac{1}{4\pi G} \partial_b V \partial_k V + \frac{1}{c^2} \Big[ 4 (\sigma_{bk} V -  \sigma_{(b} V_{k)})  \nonumber \\ 
& \qquad \quad + \frac{1}{\pi G} \Big( 2 \partial_t V_{(b} \partial_{k)} V - \partial_b V_i \partial_k V_i + 2  \partial_i V_{(b} \partial_{k)} V_i -  \frac{1}{2} \Delta (V_b V_k) \Big) \Big] \biggr\}\,,\\ 
\mathrm{TII}_L &= -\frac{(2\ell +1)}{2c^4(\ell +2)(2\ell +3)(2\ell +5)}  \epsilon_{ab\langle i_\ell} \mathop{\mathrm{FP}}_{B=0}\, \frac{\dd^3}{\dd t^3}\int \dd^3\mathbf{x}\,\vert \widetilde{\mathbf{x}}\vert^B \hat{x}_{L-1\rangle ak} \vert \mathbf{x}\vert^2 \biggl\{ \sigma_{bk} + \frac{1}{4\pi G} \partial_b V \partial_k V \biggr\}\,.
\end{align}
\end{subequations}
Using the surface terms appearing in the source moments, we have performed consistency checks on the distributional parts of the potentials~\eqref{eq:potdistr} by computing them in two different ways. As an example, the first surface term appearing in the mass quadrupole can be rewritten as
\begin{equation}\label{eq:examplesurf}
\underset{B=0}{\text{FP}}\int \dd^3 \mathbf{x} \, \vert \widetilde{\mathbf{x}}\vert^B \hat{x}^{ij}\Delta\left(\frac{V^2}{2}\right) = \underset{B=0}{\text{FP}}\int \dd^3 \mathbf{x} \, \vert \widetilde{\mathbf{x}}\vert^B\hat{x}^{ij}\left( -4\pi G \sigma V +\partial_a V \partial_a V + \frac{V \partial_t^2 V}{c^2} \right)\,,
\end{equation}
where both sides can be independently computed. In particular, the distributional part of the potential $V$ only contributes to the right-hand-side of Eq.~\eqref{eq:examplesurf} and it is crucial to take it into account to recover the correct value computed from the left-hand-side. We have performed a similar check for the potentials $V$, $V_i$, $\hat{W}$ and $\hat{R}_i$, which confirms the expressions of Eqs.~\eqref{eq:potdistr}. For the source mass quadrupole $\dI_{ij}$, as explained in~\cite{HFB20a}, such terms do not play a role as the 1PN value of $V$ only contributes to surface and compact terms, in which the distributional part does not contribute. However, for the current quadrupole $\dJ_{ij}$, we have to take them into account, as both $V$ and $V_i$ appear at 1PN in non-compact terms.

First, using the integration methods described in~\cite{Marchand:2020fpt}, we have obtained the source multipole moments in a general frame for arbitrary orbits at the order required to get the full 2.5PN waveform, as displayed in~\Cref{table:moments}. Then, we have expressed them in the CoM frame using the relations~\eqref{eq:yiviCoM} derived above. Using the convenient PN parameter
 \begin{equation}
\gamma = \frac{G\tmass}{r c^2}\,,
\end{equation}
the source mass quadrupole on quasi-circular orbits reads
\begin{align}\label{eq:Iijcirc}
& \dI_{ij}= \tmass r^2 \Biggl[ n^{\langle i} n^{j\rangle} \biggl\{\nu \biggl [1 + \Bigl(- \frac{1}{42} - \frac{13}{14} \nu \Bigr) \gamma + \Bigl(- \frac{461}{1512} -  \frac{18395}{1512} \nu  -  \frac{241}{1512} \nu^2\Bigr) \gamma^2\biggl]\nonumber\\
& \qquad + \Bigl(3 \widetilde{\mu}_{+}^{(2)} + 3 \delta\, \widetilde{\mu}_{-}^{(2)}\Bigr) \gamma^5 + \biggl [\widetilde{\mu}_{+}^{(2)} \Bigl(- \frac{3}{2} + \frac{\nu}{7} -\frac{222}{7} \nu^2\Bigr) + \delta\, \widetilde{\mu}_{-}^{(2)} \Bigl(- \frac{3}{2} -  \frac{67}{7} \nu \Bigr) + \frac{160}{3} \nu \widetilde{\sigma}_{+}^{(2)}\biggl] \gamma^6\nonumber\\
& \qquad + \biggl [\mutp \Bigl(\frac{871}{56} - \frac{1613}{168} \nu - \frac{17237}{168} \nu^2 + \frac{929}{42} \nu^3\Bigr) + \delta\, \mutm \Bigl(\frac{871}{56} + \frac{1493}{24} \nu - \frac{7201}{168} \nu^2\Bigr) + \sigmatp \Bigl(\frac{388}{9} \nu -  \frac{2504}{7} \nu^2\Bigr)\nonumber\\
& \qquad + \frac{1732}{63} \delta \nu \sigmatm\biggl] \gamma^7\biggl\} + \lambda^{\langle i} \lambda^{j\rangle} \biggl\{\nu \biggl [\Bigl(\frac{11}{21} -  \frac{11}{7} \nu \Bigr) \gamma + \Bigl(\frac{1013}{378} + \frac{299}{378} \nu -  \frac{365}{378} \nu^2\Bigr) \gamma^2\biggl] + \biggl [\widetilde{\mu}_{+}^{(2)} \Bigl(3 + \frac{104}{7} \nu -  \frac{198}{7} \nu^2\Bigr) \nn\\
& \qquad + \delta \,\widetilde{\mu}_{-}^{(2)} \Bigl(3-  \frac{38}{7} \nu \Bigr) + \frac{128}{3} \nu \widetilde{\sigma}_{+}^{(2)}\biggl] \gamma^6 + \biggl [\widetilde{\mu}_{+}^{(2)} \Bigl(- \frac{19}{2} + \frac{617}{42} \nu + \frac{5039}{42} \nu^2 + \frac{260}{21} \nu^3\Bigr) \nn\\
& \qquad + \delta \,\widetilde{\mu}_{-}^{(2)} \Bigl(- \frac{19}{2} + \frac{1291}{42} \nu -  \frac{1649}{42} \nu^2\Bigr) + \widetilde{\sigma}_{+}^{(2)} \Bigl(- \frac{64}{9} \nu -  \frac{1696}{7} \nu^2\Bigr) + \frac{2048}{63} \delta \nu \widetilde{\sigma}_{-}^{(2)}\biggl] \gamma^7\biggl\} \nn\\
& \qquad + n^{\langle i} \lambda^{j \rangle}\biggr\{\frac{48}{7}\nu^2 \gamma^{5/2} +\left[\mutp\left(-\frac{64}{5}+\frac{2336}{35}\nu+\frac{1296}{7}\nu^2 \right)+\delta\,\mutm\left(-\frac{64}{5}+\frac{288}{7}\nu \right) \right]\gamma^{15/2} \biggl\} \Biggl]\, ,
\end{align}
while the other source moments are given in Eqs.~\eqref{eq:momentcirc}. We recall that the notations are explained in~\Cref{subsec:notations}. Furthermore, as seen in~\Cref{subsec:genmoments}, we also need to compute the gauge moment W and the current dipole $\dS_i$. At leading order, they read
\begin{subequations}
\begin{align}
\dW &= \frac{1}{3} \,\underset{B=0}{\text{FP}} \int \dd^3 \mathbf{x} \, \vert \widetilde{\mathbf{x}}\vert^B x^i \sigma_i + \calO\left(\frac{1}{c^2} \right) = \frac{1}{3} \tmass \nu (x v)+ \calO\left(\frac{1}{c^2},\frac{\etidal}{c^2} \right)\,,\\
\dS_i &= \epsilon_{iab}\,\underset{B=0}{\text{FP}} \int \dd^3 \mathbf{x} \, \vert \widetilde{\mathbf{x}}\vert^B x^a \sigma^b + \calO\left(\frac{1}{c^2} \right) = \tmass \nu\, \epsilon_{iab}\,x^a v^b+ \calO\left(\frac{1}{c^2},\frac{\etidal}{c^2} \right)\,.
\end{align}
\end{subequations}
In particular, there is no tidal contribution at leading order due to the STF behaviour of the tidal tensors. On quasi-circular orbits, as $(xv)=\calO(1/c^{5})$, the gauge potential W vanishes and does not contribute to the radiative mass quadrupole.

\subsection{Treatment of the memory terms}\label{subsec:computationalmethod}

The non-linear memory contributions~\eqref{Umemory} are taken into account within the PN-MPM formalism. Such hereditary integrals are usually filed into two categories: the oscillatory integrals (or ``AC'') which take the form
\begin{equation}\label{eq:AC}
\int_{-\infty}^{T_R} \dd \tau \, x^p(\tau) e^{-\di n \phi(\tau)} = \di \frac{G \tmass}{n c^3}e^{-\di n \phi}x^{p-3/2}\,,
\end{equation}
and the non-oscillatory (or ``DC'') integrals which take the form
\begin{equation}\label{eq:DC}
\int_{-\infty}^{T_R} \dd \tau \, x^p(\tau) = \int_0^{x(T_R)} \dd y \, \frac{y^p}{\dot{x}(y)} \,.
\end{equation}
The latter is integrated using the time evolution of the frequency, $\dot{x}(y)$, as explicitely given in Eq.~\eqref{eq:xdot}. Notice that the DC integrals, after integration, reduce the PN order by 2.5PN, meaning that the first non-linear memory integral~\eqref{eq:Uijmem}, which appears at 2.5PN, actually contributes to the 0PN order after integration. Thus, in order to have a consistent waveform to $n$PN, one would have to push the PN-MPM algorithm up to $(n+2.5)$PN, which represents a difficult task. Fortunately, on quasi-circular orbits, those DC terms contribute only to the $m=0$ modes and an alternative method can be used to complete the full waveform to consistent PN order~\cite{Thorne:1980ru,Favata:2008yd}. Indeed, the memory contribution to the $(\ell,m)$ modes is given by the angular integral over the past of the source of the GW energy flux
\begin{equation}\label{eq:hlmmem_generic}
h_{\ell m }^\text{mem} = -\frac{16\pi G}{R c^4}\sqrt{\frac{(\ell-2)!}{(\ell+2)!}}\int^{T_R}_{-\infty} \dd t \int \dd \Omega \,\frac{\dd \mathcal{F}}{\dd\Omega} (\Omega) Y_{\ell m}^*(\Omega) = \int_{-\infty}^{T_R} \dd t \,\dot{h}_{\ell m }^\text{mem}\,.
\end{equation}
By expressing the flux in terms of the time derivatives of the modes, it permits to avoid using the radiative multipole moments. Then, after performing the angular integration, as explained in details in Ref.~\cite{Favata:2008yd}, Eq.~\eqref{eq:hlmmem_generic} becomes an integral over time of the following source
\begin{equation}\label{eq:hlmmemdotgen}
\dot{h}_{\ell m }^\text{mem} = -\frac{R}{c}\sqrt{\frac{(\ell-2)!}{(\ell+2)!}}\sum_{\ell'=2}^\infty\sum_{\ell''=2}^\infty \sum_{m'=-\ell'}^{\ell'}\sum_{m''=-\ell''}^{\ell''} (-1)^{m+m''} G^{2 -2 0}_{\ell' \ell'' \ell m'-m''-m}\,\dot{h}_{\ell' m'}\dot{h}^*_{\ell'' m''}\,,
\end{equation}
where the value of the angular integral $G^{s s' s''}_{\ell \ell' \ell'' m m' m''}$ is given in \textit{e.g.} Appendix A of~\cite{Favata:2008yd}. Using this approach and Eqs.~\eqref{eq:AC} and~\eqref{eq:DC}, we were able to compute the $(\ell,0)$ (for $\ell$ even) modes by integrating the AC and DC integrals to consistent PN orders.

\section{Full waveform}\label{sec:results}

We are now in a position to provide a complete description of the waveform radiated from an inspiralling, non-spinning binary system, with a precision of 2.5PN order. In~\Cref{subsec:flux&phase}, we start by recalling the tidal contributions to the radiated flux, the orbital phase, and the phase within the stationary phase approximation (SPA). By inserting the intermediate results of~\Cref{sec:intermediatecomp} in the expressions of~\Cref{subsec:genmoments}, we have computed the radiative moments at the orders given in~\Cref{table:moments}. Then, using Eqs.~\eqref{eq:hlm} and~\eqref{eq:hlmmem_generic}, we have computed the waveform amplitude modes, which constitute our main result. In~\Cref{subsec:hlm} and~\Cref{subsec:eob}, we respectively display them in the conventional PN expanded way in the form of \textit{e.g.}~\cite{Blanchet:2008je} and in a factorized form convenient for the EOB formalism. All the results of this Section are provided in the Supplemental Material~\cite{SuppMaterial}.

\subsection{Radiated flux and phase evolution}\label{subsec:flux&phase}

For binary systems on quasi-circular orbits, the GW energy flux and phase are expressed in terms of the usual gauge invariant PN parameter
\begin{equation}\label{eq:x}
x = \left(\frac{G \tmass \omega}{c^3} \right)^{2/3}\,, 
\end{equation}
tied to the orbital frequency $\omega$ and of order $\mathcal{O}(c^{-2})$. The GW energy flux is split according to
\begin{equation}
\mathcal{F} = \mathcal{F}_\text{pp}+\mathcal{F}_\text{tidal}\,,
\end{equation}
where $\mathcal{F}_\text{pp}$ is the point-particle contribution to the flux, which is currently known to 4.5PN order and can be found in Eq.~(6.11) of ~\cite{Blanchet:2023sbv}. The tidal part was derived to 2.5PN order in~\cite{HFB20a} and is given by~\footnote{In a previous version of Ref.~\cite{HFB20a}, an incorrect value of the flux was published due to the fact that they use the EoM of Ref.~\cite{HFB19} that is in a different (non-harmonic) gauge. We present here the corrected value, as well as for the phases~\cite{erratumII}.}
\begin{align}
\label{eq:Ftidal}
\mathcal{F}_{\text{tidal}} &= \dfrac{192 c^5 \nu\,x^{10}}{5G} \Biggl\{ (1+4 \nu)\widetilde{\mu}_{+}^{(2)} + \delta\, \widetilde{\mu}_{-}^{(2)} \\
& + \left[ \mutp\left(-\dfrac{22}{21} -\dfrac{1217}{168} \nu - \dfrac{155}{6}\nu^2 \right) + \delta \, \mutm \left( -\dfrac{22}{21} - \dfrac{23}{24}\nu \right) + \sigmatp\left(-\dfrac{1}{9} +\dfrac{76}{3}\nu \right) -\dfrac{1}{9}\delta\, \sigmatm  \right] x \nonumber\\
& +4\pi \left[ (1+4 \nu)\mutp + \delta\, \mutm \right] x^{3/2} \nonumber \\
& + \left[\mutp\left(\dfrac{167}{54} - \dfrac{649853}{18144} \nu + \dfrac{15923}{336}\nu^2 +\dfrac{965}{12} \nu^3 \right) + \delta\,\mutm \left( \dfrac{167}{54} + \dfrac{74783}{2016} \nu - \dfrac{2779}{144} \nu^2 \right) \right. \nonumber\\
& \qquad\left. + \sigmatp\left(-\dfrac{173}{756} + \dfrac{145}{3} \nu -208\nu^2 \right) + \delta\, \sigmatm \left(-\dfrac{173}{756} + \dfrac{1022}{27} \nu \right) + \dfrac{80}{3}\nu \widetilde{\mu}_{+}^{(3)} \right]x^2 \nonumber \\
&+ 4\pi \left[\mutp\left(-\dfrac{22}{21} - \dfrac{5053}{1344} \nu - \dfrac{2029}{48}\nu^2 \right) + \delta\,\mutm \left( - \dfrac{22}{21} - \dfrac{351}{64} \nu \right) + \sigmatp\left(-\dfrac{1}{18} + \dfrac{226}{9} \nu  \right) -\dfrac{\delta}{18} \sigmatm  \right]x^{5/2}\Biggr\}\,.\nonumber
\end{align}
The flux-balance equation~\eqref{eq:flux_balance_eq}
directly yields the time evolution of the frequency~\eqref{eq:x},
\begin{equation}\label{eq:flux_balance_xdot}
\frac{\dd x}{\dd t}= - \frac{\mathcal{F}(x)}{\dd E/ \dd x}\,,
\end{equation}
whose expression to 2.5PN order reads
\begin{align}\label{eq:xdot}
\frac{\dd x}{\dd t} &= \frac{64 c^3\nu x^5}{5G \tmass}\Biggl\{ 1+ \left( -\frac{743}{336}-\frac{11}{4}\nu \right) x + 4\pi x^{3/2} + \left( \frac{34103}{18144} + \frac{13661}{2016}\nu + \frac{59}{18}\nu^2\right)x^2 + \pi \left( -\frac{4159}{672} -\frac{189}{8}\nu\right)x^{5/2} \nn \\
& \quad + \biggl[\mutp\left(\frac{6}{\nu} +132\right) +\frac{6}{\nu}\delta \,\mutm\biggr]x^5+ \biggl[\mutp\left(\frac{19}{7\nu}+\frac{4763}{14}-661\nu\right)+\delta\,\mutm\left(\frac{19}{7\nu}+\frac{751}{4}\right) \nn \\
& \quad +\sigmatp\left(-\frac{2}{3\nu}+1384\right)-\frac{2}{3\nu}\delta\,\,\sigmatm\biggr]x^6+\frac{24\pi}{\nu}\biggl[(1+22\nu)\mutp +\delta \,\mutm\biggr]x^{13/2} \nn \\
& \quad + \biggl[ \mutp \left( \frac{21011}{252\nu} +\frac{23205233}{6048} -\frac{401215}{112}\nu +\frac{18721}{12}\nu^2 \right)+\delta\,\mutm \left( \frac{21011}{252\nu} +\frac{378769}{224} -\frac{26155}{24}\nu \right) \nn \\
& \quad + \sigmatp\left( -\frac{299}{126\nu}+\frac{55430}{9}-9120\nu\right) +\delta\,\sigmatm\left(-\frac{299}{126\nu}+\frac{7753}{3} \right) +1200\widetilde{\mu}_+^{(3)}\biggr] x^7 \nn \\
& \quad + \pi \biggl[\mutp\left(\frac{76}{7\nu} + \frac{24239}{14}-4402\nu\right) +\delta\, \mutm\left(\frac{76}{7\nu}+\frac{5139}{8}\right)+\sigmatp\left(-\frac{4}{3\nu}+\frac{16592}{3}\right)-\frac{4}{3\nu}\delta\,\sigmatm\biggr]x^{15/2}\Biggr\}\,.
\end{align}
A similar reshuffling of the energy flux balance equation~\eqref{eq:flux_balance_eq} is used to calculate the orbital phase in the time domain, namely
\begin{equation}\label{eq:flux_balance_phase}
\phi = \int \dd t \, \omega = - \frac{c^3}{G\tmass}\int  x^{3/2}\frac{\dd E/ \dd x}{\mathcal{F}(x)}\, \dd x \,.
\end{equation}
As was done for the GW energy flux, we decompose the phase as 
\begin{equation}
\phi = \phi_{\text{pp}} + \phi_{\text{tidal}}\,,
\end{equation}
and we recall the tidal contribution to 2.5PN order~\cite{HFB20a,erratumII}
\begin{align}
\phi_{\text{tidal}} &= - \dfrac{3 x^{5/2}}{16 \nu^2} \Biggl\{ (1+22 \nu)\mutp + \delta\,\mutm  \\
& + \left[ \mutp\left(\dfrac{195}{56} + \dfrac{1595}{14} \nu + \dfrac{325}{42}\nu^2 \right) + \delta \,\mutm \left( \dfrac{195}{56} + \dfrac{4415}{168}\nu \right) + \sigmatp\left(-\dfrac{5}{63} +\dfrac{3460}{21}\nu \right) -\dfrac{5}{63} \delta\,\sigmatm \right] x \nonumber\\
& -\dfrac{5\pi}{2} \left[(1+22 \nu)\mutp + \delta\, \mutm \right] x^{3/2} \nonumber\\
& + \left[\mutp\left(\dfrac{136190135}{9144576} + \dfrac{978554825}{1524096} \nu - \dfrac{281935}{2016}\nu^2 +5 \nu^3 \right) + \delta\, \mutm\left( \dfrac{136190135}{9144576} + \dfrac{213905}{864} \nu + \dfrac{1585}{432} \nu^2 \right)\right. \nonumber\\
& \qquad \left. + \sigmatp \left(-\dfrac{745}{1512} + \dfrac{1933490}{1701} \nu - \dfrac{3770}{27}\nu^2  \right) + \delta\,\sigmatm \left(-\dfrac{745}{1512} + \dfrac{19355}{81} \nu \right) + \dfrac{1000}{9}\nu \widetilde{\mu}_{+}^{(3)} \right]x^2  \nonumber\\
& + \pi \left[ \mutp\left(-\dfrac{397}{32} - \dfrac{5343}{16} \nu + \dfrac{1315}{12}\nu^2 \right) + \delta \, \mutm\left( -\dfrac{397}{32} - \dfrac{6721}{96}\nu \right) + \sigmatp\left(\dfrac{1}{3} -\dfrac{4156}{9}\nu \right) + \dfrac{\delta}{3} \sigmatm \right]x^{5/2}\Biggr\}\,.\nonumber
\end{align}
Finally, to make PN results more convenient to integrate into data analysis frameworks, it is common practice to express the phase in the Fourier domain, using the SPA~\cite{Tichy:1999pv}. The phase in the Fourier domain,
\begin{equation}
\phi^\text{SPA} = \phi_{\text{pp}}^\text{SPA} + \phi_{\text{tidal}}^\text{SPA}\,,
\end{equation}
is evaluated at the GW frequency $f$, which is twice the orbital frequency, namely $v=(\tfrac{\pi G\tmass f}{c^3})^{1/3}$. We get
\begin{align}
\phi^\text{SPA}_{\text{tidal}} =& -\dfrac{9 v^{5}}{16 \nu^2} \Biggl\{ (1+22 \nu)\mutp + \delta\, \mutm \\
& + \left[ \mutp \left(\dfrac{195}{112} + \dfrac{1595}{28} \nu + \dfrac{325}{84}\nu^2 \right) + \delta \,\mutm \left( \dfrac{195}{112} + \dfrac{4415}{336}\nu \right) + \sigmatp\left(-\dfrac{5}{126} +\dfrac{1730}{21}\nu \right) -\dfrac{5}{126} \delta \, \sigmatm \right] v^{2} \nn\\
& -\pi \left[ (1+22 \nu)\mutp + \delta \,\mutm \right]v^{3} \nn\\
& + \left[\mutp\left(\dfrac{136190135}{27433728} + \dfrac{978554825}{4572288} \nu - \dfrac{281935}{6048}\nu^2 + \dfrac{5}{3} \nu^3 \right) + \delta\,\mutm \left( \dfrac{136190135}{27433728} + \dfrac{213905}{2592} \nu + \dfrac{1585}{1296} \nu^2 \right) \right. \nn\\
& \qquad \left. + \sigmatp \left(-\dfrac{745}{4536} + \dfrac{1933490}{5103} \nu - \dfrac{3770}{81}\nu^2  \right) + \delta\, \sigmatm \left(-\dfrac{745}{4536} + \dfrac{19355}{243} \nu \right) + \dfrac{1000}{27}\nu \widetilde{\mu}_{+}^{(3)} \right]v^{4} \nn\\
& + \pi \left[ \mutp\left(-\dfrac{397}{112} - \dfrac{5343}{56} \nu + \dfrac{1315}{42}\nu^2 \right) + \delta \, \mutm \left( -\dfrac{397}{112} - \dfrac{6721}{336}\nu \right) + \sigmatp \left(\dfrac{2}{21} -\dfrac{8312}{63}\nu \right) + \dfrac{2}{21}\delta\, \sigmatm\right]v^{5}\Biggr\} \nonumber\,.
\end{align}
The expressions of $\phi_{\text{pp}}$ and $\phi^\text{SPA}_{\text{pp}}$ are given up to 4.5PN order in Eqs.~(8) and~(9) of~\cite{Blanchet:2023bwj}. Recently, the terms proportional to the tidal polarizations $\widetilde{\mu}_\pm^{(2)}$ in the GW flux and phase were computed to 2PN order using EFT techniques~\cite{Patil2024} and all results are in agreement.

\subsection{Amplitude modes: PN-expanded form}\label{subsec:hlm}

The amplitude modes defined in Eq.~\eqref{eq:hlm} can be written in terms of the phase variable $\phi = \int \dd t \,\omega$. However, in order to simplify the expressions by absorbing most of the logarithms of the orbital frequency in the oscillatory part of the modes, it is convenient to introduce the new phase variable~\cite{Blanchet:2008je},
\begin{equation}\label{psi}
\psi \equiv \phi - \frac{2G \ADM \omega}{c^3} \ln\left(\frac{\omega}{\omega_0}\right)\,,
\end{equation}
where we recall that $\ADM = \tmass + \frac{E}{c^2}$ is the ADM mass and where $\omega_0 \equiv \exp[11/12-\gamma_E]/(4 b_0)$. 
Then, the gravitational modes defined in Eq.~\eqref{eq:h} read
\begin{align}\label{eq:hlmexpl}
	h_{\ell m} = \frac{8 G \tmass \nu x}{R c^2} \,
	\sqrt{\frac{\pi}{5}}\,\left(\hat{H}^\text{pp}_{\ell m}+x^5\hat{H}^\text{tidal}_{\ell m}\right)\,e^{-\di m \psi}\,.
\end{align}
The values of the $\hat{H}_{\ell m}^\text{pp}$'s are given to 3.5PN order in~\cite{Henry:2022ccf}, while the (2,2) mode is displayed to 4PN order in~\cite{Blanchet:2023bwj} and the DC memory contribution is computed to 3PN in~\cite{Favata:2008yd}. Finally, the tidal part of the amplitude modes for the full waveform containing adiabatic tides to 2.5PN beyond the leading order read
\begin{subequations} \label{HlmExp}
\begin{align}
\hat{H}^\text{tidal}_{22}&= \frac{1}{\nu} \biggl\{\mutp(3+12\nu) + 3\delta\, \mutm + \biggl[\mutp\left(\frac{9}{2}-20\nu+\frac{45}{7}\nu^2\right) +\delta\, \mutm \left( \frac{9}{2}+\frac{125}{7}\nu\right)  +\frac{224}{3}\nu\, \sigmatp\biggr] x \nn\\
&  \quad   + 6\pi \biggl[ \mutp(1+4\nu) + \delta\, \mutm \biggr]x^{3/2}  + \biggl[\mutp\left(\frac{1403}{56}-\frac{7211}{168}\nu-\frac{19367}{168}\nu^2-\frac{274}{21}\nu^3\right) \nn  \\
& \quad  +\delta\, \mutm \left( \frac{1403}{56}+\frac{1559}{56}\nu+\frac{103}{24}\nu^2\right) + \sigmatp \left(\frac{11132}{63}\nu-\frac{6536}{63}\nu^2\right) +\frac{8084}{63}\delta\,\nu\, \sigmatm + 80\nu \,\widetilde{\mu}_+^{(3)}\biggr] x^2 \nn\\
& \quad  +\biggl[ \mutp\left( \frac{\di}{5} (64-108\nu -8640\nu^2) +\frac{\pi}{7} (63-301\nu+132\nu^2) \right) \nn\\
& \quad \quad +\delta\,\mutm\left( \frac{\di}{5} (64+20\nu)+\frac{\pi}{7} (63+229\nu) \right) +\frac{448}{3}\pi\, \nu\, \sigmatp \biggr]x^{5/2} \biggr\}\label{h22}\,,\\ 
\hat{H}^\text{tidal}_{21} &= \frac{\di}{\nu} \biggl\{ \left[ 6\delta\,\nu\,\mutp -3\nu\mutm-4\delta\,\sigmatp-4\sigmatm\right] \sqrt{x}+\left[ \delta\,\mutp\left(\frac{21}{4}\nu-\frac{55}{7}\nu^2 \right)+\mutm\left( \frac{117}{28}\nu-\frac{230}{7}\nu^2\right) \nn \right. \\
& \quad \left.+\delta\,\sigmatp\left( -\frac{32}{3} +\frac{1124}{21}\nu \right) + \sigmatm\left( -\frac{32}{3} +\frac{148}{7}\nu \right) \right]x^{3/2} + \left[ \frac{3}{2}\nu \left(2\delta\,\mutp-\mutm\right)\bigl(2\pi -\di(1+\ln 16)\bigr) \nn \right. \\
& \quad \left. + \left(\delta\,\sigmatp+\sigmatm\right)\bigl(-4\pi +2\di(1+\ln 16)\bigr) \right]x^2+ \left[ \delta\,\mutp\left(\frac{4747}{112}\nu-\frac{8753}{126}\nu^2+\frac{253}{84}\nu^3\right) \nn \right. \\
& \left. \quad  +\mutm \left( \frac{1913}{144}\nu-\frac{1493}{9}\nu^2+\frac{14473}{252}\nu^3\right) + \delta\,\sigmatp \left(-\frac{608}{21}+\frac{35669}{189}\nu-\frac{29072}{189}\nu^2\right) \nn \right. \\
& \left. \quad +\sigmatm\left(-\frac{608}{21}+\frac{4783}{27}\nu-\frac{21296}{63}\nu^2\right)  + 40\nu \,\delta\,\widetilde{\mu}_+^{(3)}\right]x^{5/2} \biggr\}\,,\\ 
\hat{H}^\text{tidal}_{20}& = \frac{15}{7\nu}\sqrt{\frac{3}{2}}\biggl\{ \nu\,\mutp + \left[\mutp\left( -\frac{1931}{42336}+\frac{32621}{10584}\nu+\frac{53}{126}\nu^2\right)+\delta\,\mutm\left( -\frac{1931}{42336} +\frac{1913}{1344}\nu \right) \right.\nn \\
& \left. \quad + \sigmatp\left(-\frac{23}{3024}+\frac{2365}{252}\nu\right)-\frac{23}{3024}\delta\, \sigmatm\right]x +\left[ \mutp\left(\frac{1315093}{24385536} +\frac{2813212739}{134120448}\nu-\frac{10461949}{798336}\nu^2 -\frac{23555}{19008}\nu^3 \right)\right. \nn\\
&\left. \quad + \delta\,\mutm\left( \frac{1315093}{24385536} +\frac{21184111}{1892352}\nu+\frac{233305}{608256}\nu^2 \right)  +\sigmatp\left(-\frac{92539}{3483648}+\frac{109223}{2688}\nu-\frac{26207}{3456}\nu^2\right)\right. \nn\\
& \left. \quad +\delta\,\sigmatm \left( -\frac{92539}{3483648}+\frac{225287}{13824}\nu\right) +\frac{65}{9}\nu\,\widetilde{\mu}_+^{(3)}\right]x^2 +\pi\left[\mutp\left(-\frac{253}{2448} -\frac{253}{204}\nu +\frac{1012}{153}\nu^2\right)\right. \nn\\
& \left. \quad + \delta\,\mutm\left(-\frac{253}{2448} -\frac{529}{2448}\nu \right) + \sigmatp\left(\frac{23}{1836}-\frac{23}{459}\nu \right)+\frac{23}{1836}\delta\,\sigmatm\right]x^{5/2} \biggr\}\,,\\ 
\hat{H}^\text{tidal}_{33} &=-\frac{27}{2}\di\sqrt{\frac{15}{14}} \biggl\{ \left(\delta\, \mutp -\mutm\right)\sqrt{x}+ \biggl[\delta\,\mutp\left(-\frac{11}{3}+\frac{5}{6}\nu\right) + \mutm \left( \frac{10}{3}-\frac{20}{3}\nu\right)  +6\delta\, \sigmatp-\frac{2}{3}\sigmatm\biggr]x^{3/2} \nn \\
&\quad + \left( \delta\,\mutp - \mutm \right) \left[ 3\pi -\frac{21}{5} \di +6\di \ln(3/2) \right]x^2 \nonumber \\ 
&\quad +\biggl[\delta\,\mutp\left(-\frac{5273}{13860}-\frac{398941}{27720}\nu-\frac{391}{990}\nu^2\right) + \mutm \left( -\frac{5683}{2772}+\frac{3723}{616}\nu-\frac{358}{99}\nu^2\right)  \nn \\
& \quad  + \delta\,\sigmatp \left(\frac{16}{9}-\frac{74}{9}\nu\right) + \sigmatm\left(\frac{32}{3}-42\nu\right) + \delta\,\widetilde{\mu}_+^{(3)}\left( \frac{5}{6\nu} + \frac{20}{3} \right)+\frac{5}{6\nu}\widetilde{\mu}_-^{(3)}\biggr]x^{5/2} \biggr\}\,,\\ 
\hat{H}^\text{tidal}_{32} &= 4\sqrt{\frac{5}{7}} \biggl\{ \left[ \mutp(3-6\nu)-\delta\,\mutm +\frac{16}{3}\sigmatp\right]x + \left[ \mutp\left( -\frac{157}{120} +\frac{17}{3} \nu +\frac{49}{12}\nu^2 \right) +\delta\,\mutm \left( \frac{109}{24} -\frac{35}{4}\nu \right) \nn \right. \\
&\quad \left. +\sigmatp\left(\frac{128}{9} -48\nu \right)+\frac{16}{3}\delta\,\sigmatm \right] x^2 \nn \\
& \quad + \left[ \mutp\left( \di\left( -\frac{79}{10} +\frac{132}{5}\nu \right) +6\pi(1-2\nu) \right) + \delta\,\mutm \left( \frac{41}{10}\di -2\pi \right) +\sigmatp \left( -16\di +\frac{32}{3}\pi \right) \right]x^{5/2}\biggr\}\,,\\ 
\hat{H}^\text{tidal}_{31} &=\frac{3\,\di}{2\sqrt{14}}  \biggl\{ \left(\delta\, \mutp -\mutm\right)\sqrt{x}+ \biggl[\delta\,\mutp\left(\frac{1}{9}-\frac{65}{18}\nu\right) + \mutm \left( \frac{4}{9}-\frac{32}{9}\nu\right)  +\frac{34}{3}\delta \,\sigmatp-6\sigmatm\biggr]x^{3/2} \nn \\
&\quad + \left[ \delta\,\mutp \left(  \pi - \frac{7}{5}\di -2\di \ln(2)\right) + \mutm \left( - \pi +\frac{7}{5} \di +2\di \ln(2)\right) \right]x^2 \nonumber \\ 
&\quad +\biggl[\delta\,\mutp\left(\frac{43753}{8316}-\frac{14347}{1848}\nu+\frac{199}{66}\nu^2\right) + \mutm \left( -\frac{923}{2772}-\frac{354367}{16632}\nu+\frac{2204}{99}\nu^2\right)  \nn \\
& \quad  + \delta\,\sigmatp \left(\frac{1144}{81}-\frac{3682}{81}\nu\right) + \sigmatm\left(\frac{632}{81}-\frac{1382}{27}\nu\right) + \delta\,\widetilde{\mu}_+^{(3)}\left( \frac{5}{6\nu} + \frac{20}{3} \right)+\frac{5}{6\nu}\widetilde{\mu}_-^{(3)}\biggr]x^{5/2} \biggr\}\,,\\
\hat{H}^\text{tidal}_{30} &= -\frac{12\,\di}{5}\sqrt{\frac{6}{7}} \biggl[ \mutp(1+4\nu) + \delta\,\mutm  \biggr]x^{5/2}\,,\\
\hat{H}^\text{tidal}_{44} &= 16\sqrt{\frac{5}{7}} \biggl\{ \left[ \mutp\left(-\frac{7}{3}+4\nu \right) +\delta\,\mutm\right]x + \left[\mutp\left( \frac{18641}{1980} -\frac{25687}{990}\nu +\frac{335}{66}\nu^2\right) +\delta\,\mutm \left( -\frac{14893}{1980}+\frac{403}{55}\nu\right) \right. \nn\\
& \quad \left. +\sigmatp\left(-\frac{368}{45}+\frac{352}{15}\nu \right)+\frac{16}{15}\delta\, \sigmatm \right]x^2 + \left[\mutp \left( \frac{\pi}{3}(-28+48\nu)+\frac{\di}{120}(2167-4772\nu)+\frac{\di}{3}(-56+96\nu)\ln 2\right) \right. \nn \\
& \left. \quad +\delta\,\mutm \left(-\frac{1193\,\di}{120}+4\pi +8\di\ln 2 \right)\right]x^{5/2}\biggr\}\,,\\
\hat{H}^\text{tidal}_{43} &= \frac{27\di}{8}\sqrt{\frac{5}{14}}\biggl\{ \left[ \delta\,\mutp(-7+8\nu)+\mutm (3-6\nu) -8\delta\,\sigmatp + 8\sigmatm\right]x^{3/2} + \left[ \delta\,\mutp\left( \frac{2366}{165} -\frac{1801}{66} \nu -\frac{24}{55}\nu^2 \right) \nn \right. \\
&\quad \left. +\mutm \left( -\frac{500}{33}+ \frac{7389}{110}\nu -\frac{8366}{165}\nu^2 \right)+\delta\,\sigmatp\left(-\frac{104}{33}+ \frac{1864}{33}\nu \right)+\sigmatm\left(-\frac{200}{11} +\frac{1096}{33}\nu \right) \right] x^{5/2}\,,\\
\hat{H}^\text{tidal}_{42} &= -\frac{2}{7}\sqrt{5} \biggl\{ \left[ \mutp\left(-\frac{7}{3}+4\nu \right) +\delta\,\mutm\right]x + \left[\mutp\left( \frac{5609}{1980} -\frac{3493}{990}\nu -\frac{601}{66}\nu^2\right) +\delta\,\mutm \left( -\frac{7477}{1980}+\frac{252}{55}\nu\right) \right. \nn\\
& \quad \left. +\sigmatp\left(-\frac{752}{45}+\frac{608}{15}\nu \right)+\frac{48}{5}\delta\, \sigmatm \right]x^2 + \left[\mutp \left( \frac{\di}{5}(42-112\nu) +\frac{\pi}{3}(-14+24\nu)\right) \right. \nn\\
& \quad \left. +\delta\,\mutm \left(-\frac{28\,\di}{5} + 2\pi \right)\right]x^{5/2}\biggr\}\,,\\ 
\hat{H}^\text{tidal}_{41} &= - \frac{\di}{56}\sqrt{\frac{5}{2}}\biggl\{ \left[ \delta\,\mutp(-7+8\nu)+\mutm (3-6\nu) -8\delta\,\sigmatp + 8\sigmatm\right]x^{3/2} + \left[ \delta\,\mutp\left( \frac{1138}{165} -\frac{1119}{110} \nu -\frac{472}{55}\nu^2 \right) \nn \right. \\
&\quad \left. +\mutm \left( -\frac{344}{33}+ \frac{5661}{110}\nu -\frac{7334}{165}\nu^2 \right)+\delta\,\sigmatp\left(-\frac{520}{33}+ \frac{760}{11}\nu \right)+\sigmatm\left(-\frac{184}{33} -\frac{152}{33}\nu \right) \right] x^{5/2}\,,\\
\hat{H}^\text{tidal}_{40} &= \frac{1}{28\nu\sqrt{2}}\biggl\{ \nu\mutp + \left[\mutp\left( -\frac{224485}{310464}-\frac{204899}{77616}\nu+\frac{17363}{462}\nu^2\right)+\delta\,\mutm\left( -\frac{224485}{310464} +\frac{25}{7392}\nu \right) \right.\nn \\
& \left. \quad + \sigmatp\left(-\frac{25}{756}+\frac{137}{21}\nu\right)-\frac{25}{756}\delta\, \sigmatm\right]x +\left[ \mutp\left(\frac{5253954437}{6974263296} +\frac{7383584291}{634023936}\nu+\frac{980528141}{27675648}\nu^2 -\frac{316183}{41184}\nu^3 \right)\right. \nn\\
&\left. \quad + \delta\,\mutm\left( \frac{5253954437}{6974263296} +\frac{6200633}{2515968}\nu+\frac{28155455}{494208}\nu^2 \right)  +\sigmatp\left(-\frac{16645}{266112}-\frac{243347}{6048}\nu+\frac{437389}{1188}\nu^2\right)\right. \nn\\
& \left. \quad +\delta\,\sigmatm \left( -\frac{16645}{266112}+\frac{382139}{57024}\nu\right) +\frac{65}{9}\nu\,\widetilde{\mu}_+^{(3)}\right]x^2 +\pi\left[\mutp\left(-\frac{13565}{8976} -\frac{13565}{748}\nu +\frac{54260}{561}\nu^2\right)\right. \nn\\
& \left. \quad + \delta\,\mutm\left(-\frac{13565}{8976} -\frac{41245}{13464}\nu \right) + \sigmatp\left(\frac{25}{459}-\frac{100}{459}\nu \right)+\frac{25}{459}\delta\,\sigmatm\right]x^{5/2} \biggr\}\,,\\ 
\hat{H}^\text{tidal}_{55} &= \frac{3125\,\di}{16\sqrt{66}}\biggl\{ \left[ \delta\,\mutp(2-2\nu) +\mutm(-1+2\nu) \right]x^{3/2}+ \left[\delta\,\mutp\left( -\frac{2191}{195}+\frac{2529}{130}\nu-\frac{229}{65}\nu^2\right) \right.  \nn \\
& \left. \quad  + \mutm\left( \frac{1087}{130}-\frac{2341}{78}\nu+\frac{207}{13}\nu^2\right) + \delta\,\sigmatp \left( \frac{56}{9} -\frac{104}{9}\nu\right)+\sigmatm\left(-\frac{8}{9} +\frac{8}{3}\nu \right)\right]x^{5/2} \biggr\} \,,\\ 
\hat{H}^\text{tidal}_{54} &= -\frac{128}{\sqrt{165}} \left[  \mutp(2-8\nu +5\nu^2) +\delta\,\mutm (-1+\nu) +\sigmatp\left(\frac{8}{3} -\frac{16}{3}\nu \right) -\frac{8}{3}\delta\,\sigmatm \right]x^2\,,\\
\hat{H}^\text{tidal}_{53} &= -\frac{27\,\di}{16}\sqrt{\frac{15}{22}}\biggl\{ \left[ \delta\,\mutp(2-2\nu) +\mutm(-1+2\nu) \right]x^{3/2}+ \left[\delta\,\mutp\left( -\frac{1159}{195}+\frac{503}{78}\nu+\frac{163}{65}\nu^2\right) \right. \nn \\
&  \left. \quad  + \mutm\left( \frac{1981}{390}-\frac{2403}{130}\nu+\frac{2137}{195}\nu^2\right) + \delta\,\sigmatp \left( \frac{40}{3} -\frac{56}{3}\nu\right)+\sigmatm\left(-8 +24\nu \right)\right]x^{5/2}  \biggr\} \,,\\ 
\hat{H}^\text{tidal}_{52} &= \frac{8}{3\sqrt{55}} \left[  \mutp(2-8\nu +5\nu^2) +\delta\,\mutm (-1+\nu) +\sigmatp\left(\frac{8}{3} -\frac{16}{3}\nu \right) -\frac{8}{3}\delta\,\sigmatm \right]x^2\,,\\
\hat{H}^\text{tidal}_{51} &= \frac{\di}{48}\sqrt{\frac{5}{77}}\biggl\{ \left[ \delta\,\mutp(2-2\nu) +\mutm(-1+2\nu) \right]x^{3/2}+ \left[\delta\,\mutp\left( -\frac{643}{195}-\frac{7}{130}\nu+\frac{359}{65}\nu^2\right) \right. \nn \\
& \left. \quad  + \mutm\left( \frac{447}{130}-\frac{4961}{390}\nu+\frac{551}{65}\nu^2\right) + \delta\,\sigmatp \left( \frac{152}{9} -\frac{200}{9}\nu\right)+\sigmatm\left(-\frac{104}{9} +\frac{104}{3}\nu \right)\right]x^{5/2}  \biggr\}\,,\\
\hat{H}^\text{tidal}_{50} &= 0\,,\\ 
\hat{H}^\text{tidal}_{66}&= -\frac{486}{\sqrt{143}} \left[ \mutp \left( -\frac{9}{5}+7\nu-4\nu^2\right)+\delta\,\mutm(1-\nu)  \right]x^2 \,,\\ 
\hat{H}^\text{tidal}_{65}&= \frac{15625\,\di}{18\sqrt{429}} \left[ \delta\,\mutp\left( \frac{27}{40} -\frac{39}{20}\nu +\frac{9}{10}\nu^2 \right) +\mutm \left( -\frac{3}{8} +\frac{3}{2}\nu -\frac{3}{4}\nu^2\right) +\delta\,\sigmatp (1-\nu)+ \sigmatm(-1+3\nu)  \right]x^{5/2}\,,\\ 
\hat{H}^\text{tidal}_{64}&= \frac{128}{11}\sqrt{\frac{2}{39}} \left[ \mutp \left( -\frac{9}{5}+7\nu-4\nu^2\right)+\delta\,\mutm(1-\nu)  \right]x^2 \,,\\ 
\hat{H}^\text{tidal}_{63}&= -\frac{81\,\di}{22}\sqrt{\frac{5}{13}} \left[ \delta\,\mutp\left( \frac{27}{40} -\frac{39}{20}\nu +\frac{9}{10}\nu^2 \right) +\mutm \left( -\frac{3}{8} +\frac{3}{2}\nu -\frac{3}{4}\nu^2\right) +\delta\,\sigmatp (1-\nu)+ \sigmatm(-1+3\nu)  \right]x^{5/2}\,,\\ 
\hat{H}^\text{tidal}_{62} &= -\frac{2}{33}\sqrt{\frac{5}{13}}\left[ \mutp \left( -\frac{9}{5}+7\nu-4\nu^2\right)+\delta\,\mutm(1-\nu)  \right]x^2\,,\\ 
\hat{H}^\text{tidal}_{61} &= \frac{5\,\di}{297\sqrt{26}} \left[ \delta\,\mutp\left( \frac{27}{40} -\frac{39}{20}\nu +\frac{9}{10}\nu^2 \right) +\mutm \left( -\frac{3}{8} +\frac{3}{2}\nu -\frac{3}{4}\nu^2\right) +\delta\,\sigmatp (1-\nu)+ \sigmatm(-1+3\nu)  \right]x^{5/2}\,,\\ 
\hat{H}^\text{tidal}_{60} &= -\frac{4835}{827904\nu\sqrt{273}}\biggl\{ \left[\mutp\left( 1+\frac{10196}{967}\nu-\frac{57792}{967}\nu^2\right)+\delta\,\mutm\left( 1 +\frac{1806}{967}\nu \right)\right]x \nn\\
&\quad +\left[ \mutp\left(-\frac{16571729}{2784960} +\frac{101920643}{2784960}\nu+\frac{7852243}{46416}\nu^2 -\frac{791987}{967}\nu^3 \right)  \right. \nn\\
& \left. \quad + \delta\,\mutm\left( -\frac{16571729}{2784960} +\frac{934241}{30944}\nu-\frac{140777}{7736}\nu^2 \right) +\sigmatp\left(-\frac{4949}{69624}+\frac{88088}{967}\nu-\frac{356230}{967}\nu^2\right) \right. \nn\\
& \left. \quad +\delta\,\sigmatm \left( -\frac{4949}{69624}+\frac{428995}{5802}\nu\right) \right]x^2 +\pi\left[\mutp\left(\frac{30576}{16439} +\frac{366912}{16439}\nu -\frac{1956864}{16439}\nu^2\right) \right. \nn\\
& \left. \quad + \delta\,\mutm\left(\frac{30576}{16439} +\frac{61152}{16439}\nu \right)\right]x^{5/2} \biggr\}\,,\\ 
\hat{H}^\text{tidal}_{77} &= \frac{117649\,\di}{160}\sqrt{\frac{7}{858}}\left[ \delta\,\mutp \left( -\frac{5}{3}+\frac{14}{3}\nu-2\nu^2\right)+\mutm(1-4\nu+2\nu^2)  \right]x^{5/2} \,,\\
\hat{H}^\text{tidal}_{75} &= -\frac{15625\,\di}{416\sqrt{66}}\left[ \delta\,\mutp \left( -\frac{5}{3}+\frac{14}{3}\nu-2\nu^2\right)+\mutm(1-4\nu+2\nu^2)  \right]x^{5/2} \,,\\
\hat{H}^\text{tidal}_{73} &= \frac{2187\,\di}{22880}\sqrt{\frac{3}{2}}\left[ \delta\,\mutp \left( -\frac{5}{3}+\frac{14}{3}\nu-2\nu^2\right)+\mutm(1-4\nu+2\nu^2)  \right]x^{5/2} \,,\\
\hat{H}^\text{tidal}_{71} &= -\frac{\di}{13728\sqrt{2}}\left[ \delta\,\mutp \left( -\frac{5}{3}+\frac{14}{3}\nu-2\nu^2\right)+\mutm(1-4\nu+2\nu^2)  \right]x^{5/2} \,,
\end{align}
\end{subequations}
where the remainder, $\mathcal{O}(\etidal\, c^{-6})$, has been omitted. One can deduce from the modes their negative $m$ correspondance using $\hat{H}_{\ell, -m}=(-1)^\ell \hat{H}_{\ell m}^*$ where the star refers to the complex conjugate operation.

In~\cite{Gamba:2023mww}, the authors derived the amplitude modes $(2,2)$ at 2PN and $(2,1)$, $(3,3)$ and $(3,1)$ to 1PN order, from the flux modes given in Eqs.~(5.4) of~\cite{HFB20a}, and incorporate these tidal corrections into the \texttt{TEOBResumS} model in an additive form. As already explained, there was a mistake in the $(2,2)$ flux mode at 2PN of a previous version of~\cite{HFB20a}, which lead to a wrong value of this amplitude mode in~\cite{Gamba:2023mww}. The error in~\cite{HFB20a} has now been corrected in~\cite{erratumII}.
Finally, the 2PN part of the (2,2) mode~\eqref{h22} is in agreement with the result derived in~\cite{Patil2024} via an EFT method.


\subsection{Amplitude modes: Effective-One-Body factorized form}\label{subsec:eob} 

In EOB waveform models, there is a freedom on the choice of resumming the waveform modes. Historically~\cite{Damour:2007xr,Damour:2007yf,Damour:2008gu,Pan:2010hz}, the choice on the form of the modes converged to the following resummation in order to have lower mismatch with numerical relativity simulations,
\begin{equation}
\label{hlmFact}
h_{\ell m}^\text{F} = h_{\ell m}^\text{N} \,\hat{S}_\text{eff}\, T_{\ell m}\, f_{\ell m}\, e^{\di \delta_{\ell m}}\,,
\end{equation}
which is constructed to match the modes in the PN-expanded form of Eqs.~\eqref{eq:hlmexpl}--\eqref{HlmExp}. The modes are factorized in five blocks. The first one, $h_{\ell m}^\text{N}$, is the leading PN order contribution. It is known analytically for each $\ell$ and $m$~\cite{Thorne:1980ru,Kidder:2007rt,Pompili:2023tna}. The second factor is the effective source term $\hat{S}_\text{eff}$ which, depending on the parity of $\ell+m$, is either the effective energy $H_\text{eff}$ or the norm of the conserved angular momentum $J$ of the system
\begin{equation}
\hat{S}_\text{eff} =
\begin{cases}
     \frac{H_\text{eff}(x)}{\tmass c^2\nu} & \text{for } \ell + m \text{ even} \\
     \frac{c\sqrt{x}}{G\tmass^2\nu}J(x) & \text{for } \ell + m \text{ odd}
    \end{cases},
\end{equation}
where $H_\text{eff}$ is related to the ADM mass $\ADM$ via the EOB energy map $H_\text{EOB}/c^2 = \ADM = \tmass \sqrt{1+2\nu \left(\tfrac{H_\text{eff}}{\tmass c^2\nu} - 1\right)}$ and $J$ is given to 2PN including tidal effects in Eqs. (6.6) of~\cite{HFB19}. The third factor $T_{\ell m}$, similarly to the phase redefinition \eqref{psi}, absorbs the ``leading logarithms'' induced by tail effects. It is given by
\begin{equation}
T_{\ell m} = \frac{\Gamma\left(\ell + 1 - 2 \di \hat{k}\right)}{\Gamma (\ell + 1)} e^{\pi \hat{k}} e^{2\di \hat{k} \ln (2m\omega b_0)},
\end{equation}
where $\Gamma$ is the Euler gamma function, $\hat{k}\equiv m \tfrac{G\ADM\omega}{c^3}$ and the constant $b_0$ is defined in the tail integrals~\eqref{eq:tails}. The remaining part of the factorized modes is expressed as an amplitude $f_{\ell m}$ and a residual phase $\delta_{\ell m}$, which are computed such that the expansion of $h_{\ell m}^\text{F}$ agrees with the PN-expanded modes in Eq.~\eqref{eq:hlmexpl}. We further split the point-particle and tidal contributions as 
\begin{equation}\label{eq:flm}
f_{\ell m} = f_{\ell m}^\text{pp}+x^5 f_{\ell m}^\text{tidal}\,,
\end{equation}
where the point-particle part to 3.5PN is available in~\cite{Henry:2022ccf}. The tidal part of the (2,2) mode to 2.5PN beyond leading order read
\begin{align}\label{eq:f22}
f^\text{tidal}_{22}&=\frac{1}{\nu} \Biggl\{ \mutp(3+12\nu) + 3\delta\,\mutm + \biggl[\mutp\left(6-23\nu+\frac{45}{7}\nu^2\right) +\delta\,\mutm \left( 6+\frac{125}{7}\nu\right) +\frac{224}{3}\nu \sigmatp\biggr] x  \nn \\
& \quad + \biggl[\mutp\left(\frac{377}{14}-\frac{11969}{168}\nu-\frac{17615}{168}\nu^2-\frac{274}{21}\nu^3\right) +\delta\, \mutm \left( \frac{377}{14}+\frac{1261}{56}\nu+\frac{103}{24}\nu^2\right) +\sigmatp\left(\frac{7940}{63}\nu-\frac{6536}{63}\nu^2\right) \nn\\
&  \quad \quad +\frac{8084}{63}\delta\,\nu\, \sigmatm + 80\nu\, \widetilde{\mu}_+^{(3)}\biggr] x^2\Biggr\} + \calO\left( \frac{\etidal}{c^6}\right)\,,
\end{align}
and the other modes can be found in Eqs.~\eqref{eq:flmvalues}.
Note the presence of $1/\delta$ factors within some modes in Eqs.~\eqref{eq:flmvalues}, which is problematic for numerical codes in the case of identical bodies. However, this is an artifact of the EOB factorization because the PN modes $h_{\ell m}$ for odd $\ell+m$ vanish in such a case. 
Next, we can further resum the amplitude terms by introducing $\rho_{\ell m} = (f_{\ell m})^{1/\ell}$ which improves the agreement with numerical-relativity waveforms \cite{Damour:2008gu,Pan:2010hz}. Still splitting $\rho_{\ell m}$ in its point-particle and tidal part,
\begin{equation}
\rho_{\ell m} = \rho_{\ell m}^\text{pp}+x^5 \rho_{\ell m}^\text{tidal}\,,
\end{equation}
the (2,2) mode is given by
\begin{align}\label{eq:rho22}
\rho^\text{tidal}_{22}&=\frac{3}{2\nu} \Biggl\{ \mutp(1+4\nu) + \delta\,\mutm + \biggl[\mutp\left(\frac{127}{42}-\frac{355}{84}\nu-\frac{10}{21}\nu^2\right) +\delta\,\mutm \left( \frac{127}{42}+\frac{445}{84}\nu\right) +\frac{224}{9}\nu \sigmatp\biggr] x  \nn \\
& \quad + \biggl[\mutp\left(\frac{148325}{10584}-\frac{438733}{21168}\nu-\frac{379511}{14112}\nu^2-\frac{62315}{10584}\nu^3\right) +\delta\, \mutm \left( \frac{148325}{10584}+\frac{264811}{21168}\nu-\frac{105869}{42336}\nu^2\right)  \nn\\
&  \quad \quad +\sigmatp\left(\frac{4252}{63}\nu-\frac{9616}{189}\nu^2\right) +\frac{8084}{189}\delta\,\nu\, \sigmatm + \frac{80}{3}\nu\, \widetilde{\mu}_+^{(3)}\biggr] x^2\Biggr\} + \calO\left( \frac{\etidal}{c^6}\right)\,.
\end{align}
and the other modes can be found in Eqs.~\eqref{eq:rholm}.
Finally, the dephasing parameters $\delta_{\ell m}$ read
\begin{subequations}
\begin{align}\label{delta22}
\delta_{22} &= \frac{7}{3}x^{3/2} -\frac{151}{6}\nu x^{5/2} + \frac{64}{5\nu}\left[ \mutp\left(1+ \frac{63}{16}\nu-\frac{7095}{64}\nu^2\right)+\delta\,\mutm\left( 1+\frac{95}{16}\nu \right)\right]x^{15/2} + \calO\left( \frac{1}{c^6},\frac{\etidal}{c^6}\right)\,,\\
\delta_{32} &= \frac{10+33\nu}{15(1-3\nu)}x^{3/2} + \frac{1}{5(1-3\nu)^2}\biggl[\mutp(66-450\nu) +\delta\, \mutm(66+54\nu)-1344\nu\sigmatp  \biggr] x^{13/2} + \calO\left( \frac{1}{c^5},\frac{\etidal}{c^5}\right) \,,\\
\delta_{44} &= \frac{112+219\nu}{120(1-3\nu)}x^{3/2} + \frac{111}{4(1-3\nu)^2}\biggl[ \mutp(1-6\nu) +\delta\,\mutm\biggr]x^{13/2} + \calO\left( \frac{1}{c^5},\frac{\etidal}{c^5}\right)\,,\\
\delta_{42} &= \frac{7+42\nu}{15(1-3\nu)}x^{3/2} + \frac{126}{5(1-3\nu)^2}\biggl[ \mutp(1-6\nu)  +\delta\,\mutm\biggr]x^{13/2} + \calO\left( \frac{1}{c^5},\frac{\etidal}{c^5}\right)\,,
\end{align}
\end{subequations}
while the other values of $\delta_{\ell m}$ do not contain tidal corrections to the considered PN order, especially the (2,1), (3,3) and (3,1) modes. Note that there is no contribution at the tidal leading 1.5PN relative order in Eq.~\eqref{delta22}. All these expressions are available in a \emph{Mathematica} file in the Supplemental Materials~\cite{SuppMaterial} including all point-particle and tidal effects for the 2.5PN waveform.

Alternatively, following Refs.~\cite{Damour:2007yf,Damour:2008gu,Pompili:2023tna}, one can express the $\delta_{\ell m}$ in terms of the variable $y \equiv (\omega H_\text{EOB})^{2/3}$. 
In our notations, this variable reads\footnote{The definition of $y$ matches the one of Ref.~\cite{Damour:2008gu} while $x$ is written $\bar{y}$.}
\begin{equation}
y = \left(\frac{G\ADM\omega}{c^3}\right)^{2/3} = \left(\frac{\ADM}{\tmass}\right)^{2/3}x\,.
\end{equation}
The expressions of $\delta_{32}$, $\delta_{44}$ and $\delta_{42}$ remain unchanged when replacing $x$ by $y$ while the expression of the (2,2) dephasing becomes
\begin{equation}\label{eq:delta22}
\delta_{22} = \frac{7}{3}y^{3/2} - 24\nu x^{5/2} + \frac{64}{5\nu}\left[ \mutp\left(1+ \frac{63}{16}\nu-\frac{225}{2}\nu^2\right)+\delta\,\mutm\left( 1+\frac{95}{16}\nu \right)\right]x^{15/2} + \calO\left( \frac{1}{c^6},\frac{\etidal}{c^6}\right)\,.
\end{equation}
Note that the 2.5PN coefficients are written in terms of $x$. Since both $x$ and $y$ are equal at leading order, one can choose freely to replace $x$ by $y$ in~\eqref{eq:delta22} for those 2.5PN order terms. If one were to include higher PN orders in a given model, notably for the point-particle part, one should care about the difference between those two variables as the higher PN orders would be affected.

\section{Summary}

This work concludes the project previously undertaken aiming at describing the effects of adiabatic tides on the orbital dynamics of non-spinning compact binaries during their inspiral phase, as well as their impact on the emitted gravitational waves. All calculations were performed within the PN formalism, achieving a precision of 2.5PN beyond the leading order, consistently applied to both the dynamical and radiative sectors. Specifically, a Fokker Lagrangian approach was used in~\cite{HFB19} to derive the conservative EoM for the system and its conserved integrals in harmonic coordinates to 2PN order. Subsequently, the tidal contribution to the gravitational flux and phase for a binary system on quasi-circular orbits was computed to 2.5PN order in~\cite{HFB20a,erratumII}, utilizing a GW generation formalism based on the PN-MPM framework, which is valid for any matter source undergoing gravitational interaction. 

In the present paper, we have extended the work of~\cite{HFB20a} by including adiabatic tidal corrections to the waveform amplitude modes, up to 2.5PN order, thus achieving the same accuracy as for the gravitational phase. To this end, we derived the multipole moments to high PN orders, specifically the mass quadrupole to 2.5PN order, and the mass octupole and current quadrupole to 2PN order. We also accounted for tidal corrections to the tail, instantaneous and memory parts of the radiative moments, which arise from non-linear effects in the propagation of GW. Next, by decomposing the gravitational wave polarizations in terms of spherical harmonics and inserting the expressions for the radiative moments, we derived all the waveform amplitude modes $h_{\ell m}$ with $\ell \leq 7$ and $\lvert m \rvert \leq 7$ to 2.5PN order. This ended the computation of the full waveform amplitude at the same 2.5PN order. These modes are displayed both in a PN-expanded form and in a factorized form convenient for direct inclusion into EOB models.

One can think at different directions to improve such waveform models. For instance, one could describe adiabatic tides for extended bodies on eccentric orbits, building on the previous investigation of the effects of eccentricity on the point-particle waveform to 3PN order~\cite{Arun:2007rg,Arun:2007sg,Arun:2009mc,Mishra:2015bqa,Boetzel:2019nfw,Ebersold:2019kdc}. Additionally, one could also go beyond the adiabatic tide approximation and consider dynamical tidal effects~\cite{Steinhoff:2016rfi,Mandal:2023hqa,Pitre:2023xsr,HegadeKR:2024agt}, particularly in the radiative sector. Lastly, one could further investigate dissipative tidal effects
~\cite{Chia:2020yla,Jones:2023ugm,Jakobsen:2023pvx,Ripley:2023lsq,HegadeKR:2024slr,Bautista:2024emt}.

\section*{Acknowledgments}

Q.H. thanks Raj Patil for useful discussions and exchanges that led to correct a mistake in previous works. E.D. is grateful for the hospitality of the Max Planck Institute for Gravitational Physics (Albert Einstein Institute, AEI) in Potsdam, where part of this work was carried out. E. D. acknowledges financial support from the SALTO exchange program between the Centre National de la Recherche Scientifique (CNRS) and the Max-Planck-Gesellschaft (MPG). This work was supported by the Universitat de les Illes Balears (UIB); the Spanish Agencia Estatal de Investigación grants PID2022-138626NB-I00, RED2022-134204-E, RED2022-134411-T, funded by MICIU/AEI/10.13039/501100011033 and the ERDF/EU; and the Comunitat Autònoma de les Illes Balears through the Conselleria d'Educació i Universitats with funds from the European Union - NextGenerationEU/PRTR-C17.I1 (SINCO2022/6719).
L.B. acknowledges financial support from the ANR PRoGRAM project, grant ANR-21-CE31-0003-001 and the EU Horizon 2020 Research and Innovation Programme under the Marie Sklodowska-Curie Grant Agreement no. 101007855.
This research was supported in part by Perimeter Institute for Theoretical Physics. Research at Perimeter Institute is supported in part by the Government of Canada through the Department of Innovation, Science and Economic Development and by the Province of Ontario through the Ministry of Colleges and Universities.

\appendix
\section{Corrected expressions of the conservative sector}\label{sec:correctedConservative}

In this Appendix, we provide the corrected expressions of the relevant quantities published in~\cite{HFB19} in the same gauge as the source moments. We define $a_n = (an)$, $a_v = (av)$ and $\Lambda=\tfrac{G^2\tmass^2\nu}{r^6}$. The reduced Lagrangian in the CoM frame, leading to the same EoM up to 2PN derived in Section~\ref{subsec:RR} reads
\begin{align}\label{eq:LCoMtidal}
\frac{L_{\text{tidal}}}{\Lambda} =& \, 3 \mu_{+}^{(2)} + \frac{1}{c^{2}} \Biggl\{ \left[\mu_{+}^{(2)} \left(\frac{27}{2} + 9 \nu \right) + \frac{45}{2} \delta\,\mu_{-}^{(2)} - 24 \sigma_{+}^{(2)}\right](nv)^2  \nn \\
& + \left[\mu_{+}^{(2)} \left(\frac{15}{4} + \frac{3}{2} \nu \right) -  \frac{15}{4} \delta\,\mu_{-}^{(2)} + 24 \sigma_{+}^{(2)}\right] v^{2}  + \frac{G m}{r} \left(- \frac{27}{2} \mu_{+}^{(2)} + \frac{15}{2} \delta\,\mu_{-}^{(2)}\right)\Biggr\} \nn\\
&  + \frac{1}{c^{4}} \Biggl\{ r \biggl[\left(\mu_{+}^{(2)} \left(21 -  \frac{45}{2} \nu \right) + \delta\,\mu_{-}^{(2)} \left(21 -  \frac{9}{2} \nu \right)- 48 \nu \sigma_{+}^{(2)} \right) a_{v} (nv) \nn\\
& + \left(\mu_{+}^{(2)} \Bigl(-60 + 18 \nu \Bigr) + \delta\,\mu_{-}^{(2)} \Bigl(-60 + 18 \nu \Bigr) + \sigma_{+}^{(2)} \Bigl(-16 + 48 \nu \Bigr) - 16 \delta\,\sigma_{-}^{(2)}\right) a_{n} (nv)^2\nn\\
& + \left(\mu_{+}^{(2)} \left(\frac{39}{2} -  \frac{27}{4} \nu \right) + \delta\,\mu_{-}^{(2)} \left(\frac{39}{2} -  \frac{9}{4} \nu \right) + 16 \sigma_{+}^{(2)} + 16 \delta\,\sigma_{-}^{(2)}\right) a_{n} v^{2}\biggl] \nn \\
& + \biggl[\mu_{+}^{(2)} \Bigl(36 - 72 \nu + 18 \nu^2\Bigr) + \delta\,\mu_{-}^{(2)} \Bigl(27 - 18 \nu \Bigr) + \sigma_{+}^{(2)} \Bigl(72 - 96 \nu \Bigr) + 48 \delta\,\sigma_{-}^{(2)}\biggl] (nv)^4 \nn\\
& + \biggl[\mu_{+}^{(2)} \Bigl(- \frac{189}{4} + 72 \nu -  \frac{45}{2} \nu^2\Bigr) + \delta\,\mu_{-}^{(2)} \Bigl(- \frac{99}{4} -  \frac{27}{2} \nu \Bigr) + \sigma_{+}^{(2)} \Bigl(-114 + 132 \nu \Bigr) - 54 \delta\,\sigma_{-}^{(2)}\biggl] (nv)^2 v^{2} \nn\\
& + \biggl[\mu_{+}^{(2)} \left(\frac{249}{16} - 12 \nu -  \frac{27}{8} \nu^2\right) + \delta\,\mu_{-}^{(2)} \left(\frac{39}{16} + \frac{27}{8} \nu \right)+ \sigma_{+}^{(2)} \Bigl(42 - 36 \nu \Bigr)
 + 6 \delta\,\sigma_{-}^{(2)}\biggl] v^{4} \nn \\
&  + \frac{G m}{r} \biggl[\left(\mu_{+}^{(2)} \left(- \frac{165}{2} + \frac{355}{2} \nu + 39 \nu^2\right) + \delta\,\mu_{-}^{(2)} \left(- \frac{219}{2} + \frac{135}{2} \nu \right) + 28 \sigma_{+}^{(2)} - 44 \delta\,\sigma_{-}^{(2)}\right) (nv)^2\nn\\
& + \left(\mu_{+}^{(2)} \left(\frac{99}{4} - 41 \nu + 3 \nu^2\right) + \frac{189}{4} \delta\,\mu_{-}^{(2)} - 28 \sigma_{+}^{(2)} + 44 \delta\,\sigma_{-}^{(2)}\right) v^{2}\biggl]\nonumber\\
& + \frac{G^2 m^2}{r^2} \left[\mu_{+}^{(2)} \left(\frac{1083}{28} + \frac{3119}{28} \nu \right) -  \frac{1227}{28} \delta\,\mu_{-}^{(2)}\right] \Biggr\} + \frac{15}{r^2}\mu_{+}^{(3)}  + \mathcal{O}\left(
  \frac{\epsilon_\text{tidal}}{c^{6}} \right)\,.
\end{align}
The tidal part of the conserved energy associated to this Lagrangian reads
\begin{align}\label{eq:Etidal}
\frac{E_\text{tidal}}{\Lambda} =& - 3 \mu_{+}^{(2)} + \frac{1}{c^2} \Biggl\{ \left[\mu_{+}^{(2)} \left(\frac{27}{2} + 9\nu \right) + \frac{45}{2} \delta \, \mu_{-}^{(2)} -24 \sigma_{+}^{(2)} \right](nv)^{2} \nn\\
&  + \left[ \mu_{+}^{(2)} \left(\frac{15}{4} + \frac{3}{2}\nu \right) - \frac{15}{4} \delta \, \mu_{-}^{(2)} + 24 \sigma_{+}^{(2)} \right] v^{2}  + \frac{G \tmass}{r} \left[ \frac{27}{2} \mu_{+}^{(2)} - \frac{15}{2} \delta \, \mu_{-}^{(2)} \right] \Biggr\} \nn \\ 
& + \frac{1}{c^{4}} \Biggl\{ \left[ \mu_{+}^{(2)} \left( -372-72 \nu +54 \nu^{2} \right) + \delta \, \mu_{-}^{(2)} \left(-399 +90 \nu \right) + \sigma_{+}^{(2)} \left(88+96\nu \right) + 16 \delta \, \sigma_{-}^{(2)}\right](nv)^{4}  \nn \\
& \quad + \left[ \mu_{+}^{(2)} \left( \frac{1125}{4}-\frac{27}{2} \nu -\frac{135}{2} \nu^{2} \right)+ \delta \, \mu_{-}^{(2)}\left(\frac{1395}{4} -135 \nu \right)  + \sigma_{+}^{(2)}\left(-198-36\nu \right) -18 \delta \, \sigma_{-}^{(2)} \right] (nv)^{2}v^{2} \nn \\
&  \quad + \left[ \mu_{+}^{(2)} \left( \frac{99}{16}- \frac{27}{4} \nu -\frac{81}{8} \nu^{2} \right)+ \delta \, \mu_{-}^{(2)} \left(-\frac{531}{16} +\frac{135}{8} \nu \right) + \sigma_{+}^{(2)}\left(110-60\nu \right) + 2 \delta \, \sigma_{-}^{(2)}\right]v^4 \nn \\
&  \quad\left. + \frac{G \tmass}{r} \left[ \left(\mu_{+}^{(2)} \left( -\frac{129}{2}+\frac{499}{2} \nu + 39 \nu^{2} \right) +\delta \, \mu_{-}^{(2)} \left(-\frac{183}{2} +\frac{135}{2} \nu \right)  + \sigma_{+}^{(2)} \left(60+48\nu \right) - 12 \delta \, \sigma_{-}^{(2)}\right) (nv)^{2} \right. \right. \nn \\
& \quad \left.  + \left(\mu_{+}^{(2)} \left( \frac{27}{4}- 113 \nu +3 \nu^{2} \right) +\frac{117}{4}  \delta \, \mu_{-}^{(2)} + \sigma_{+}^{(2)} \left(-60-48\nu \right) + 12 \delta \, \sigma_{-}^{(2)}\right) v^{2} \right]  \nn \\
& \quad  + \frac{G^2 \tmass^2}{r^2} \left[ \mu_{+}^{(2)} \left( -\frac{1083}{28} - \frac{3119}{28} \nu \right) + \frac{1227}{28}\delta \, \mu_{-}^{(2)}\right] \Biggr\}  - \frac{15}{r^2}\mu_{+}^{(3)} + \mathcal{O} \left(\frac{\epsilon_\text{tidal}}{c^{6}} \right)\,,
\end{align}
and the tidal part of the angular momentum is given by
\begin{align}\label{eq:Jtidal}
\frac{J^{i}_\text{tidal}}{\Lambda} =& \,\epsilon_{ijk}x^j v^k \Biggl\{\frac{1}{c^2} \left[\mu_{+}^{(2)} \left(\frac{15}{2} + 3 \nu \right) -  \frac{15}{2} \delta\,\mu_{-}^{(2)} + 48 \sigma_{+}^{(2)}\right] \nn \\
& + \frac{1}{c^4} \Biggl[\left(\mu_{+}^{(2)} \left(\frac{303}{2} - 27 \nu - 45 \nu^2\right) + \delta\,\mu_{-}^{(2)} \left(\frac{393}{2} - 90 \nu \right) + \sigma_{+}^{(2)} \left(-196 - 120 \nu \right) - 76 \delta\,\sigma_{-}^{(2)}\right) (nv)^2 \nn \\
& \qquad + \left(\mu_{+}^{(2)} \left(\frac{9}{4} - 12 \nu -  \frac{27}{2} \nu^2\right) + \delta\,\mu_{-}^{(2)} \left(- \frac{201}{4} + \frac{45}{2} \nu \right) + \sigma_{+}^{(2)} \left(136 - 96 \nu \right) - 8 \delta\,\sigma_{-}^{(2)}\right) v^{2}\nn \\
& \qquad+ \frac{G \tmass}{r} \left(\mu_{+}^{(2)} \left(\frac{63}{2} - 154 \nu + 6 \nu^2\right) + \frac{153}{2} \delta\,\mu_{-}^{(2)} + \sigma_{+}^{(2)} \left(-88 - 48 \nu \right) + 56 \delta\,\sigma_{-}^{(2)}\right)\Biggl]\Biggr\} + \mathcal{O}\left(
  \frac{\epsilon_\text{tidal}}{c^{6}} \right)\,.
\end{align}
On quasi-circular orbits, the square of the orbital frequency $\omega^2 = (\omega^2)_\text{pp} + (\omega^2)_\text{tidal}$, expressed in terms of the separation, reads
\begin{subequations}\label{eq:omega2}
\begin{align}
(\omega^{2})_\text{pp} &= \frac{G m}{r^{3}}\left[ 1 + (-3+\nu)\gamma + \left( 6 + \frac{41}{4} \nu + \nu^{2} \right) \gamma^{2}\right] + \mathcal{O}\left( \frac{1}{c^{6}} \right)\,,\\ 
(\omega^{2})_\text{tidal} &= \frac{G m}{r^{3}}\Biggl\{ 18\,\mutp \gamma^{5} + \left[ \mutp\left(
    -\frac{249}{2} + 51\nu \right) + \frac{75}{2}\delta \, \mutm +96\sigmatp \right] \gamma^{6} \nn\\ 
& \qquad\quad + \left[ \mutp\left( \frac{35325}{56} +
\frac{2976}{7} \nu + 54\nu^{2}\right) +\delta \, \mutm \left( -\frac{11043}{56} +90\nu \right)  + \sigmatp \bigl(-616+264\nu\bigr)  \right. \nn\\
& \qquad\qquad \left. + 200 \delta \, \widetilde{\sigma}_{-}^{(2)} + 120\,\widetilde{\mu}_{+}^{(3)}\right]\gamma^{7}\Biggr\} +\mathcal{O}\left( \frac{\epsilon_\text{tidal}}{c^{6}} \right)\,.
\end{align}
\end{subequations}
It can be related to the PN parameter $x$ defined in Eq.~\eqref{eq:x} through
\begin{subequations}\label{eq:gammaofx}
\begin{align}
\gamma_\text{pp} &= x \left[ 1 + \left(1-\frac{\nu}{3}\right)x + \left( 1 - \frac{65}{12} \nu \right)x^{2} \right] + \mathcal{O}\left(\frac{1}{c^{6}} \right) \,,\\
\gamma_\text{tidal} &= x \Biggl\{ -6\mutp x^{5} + \left[ \mutp \left( -\frac{37}{2} +3 \nu \right) - \frac{25}{2}\delta \,\mutm -32\sigmatp \right] x^{6}
\nn\\ 
& \qquad\quad + \left[ \mutp\left( -\frac{4691}{56} +\frac{1105}{21} \nu + 15\nu^2\right) + \delta\,\mutm \left( -\frac{4019}{56} +\frac{95}{6}\nu \right) + \sigmatp \left(-\frac{440}{3}+\frac{88}{3}\nu\right) \right. \nn\\
& \qquad\qquad \left.  - \frac{200}{3} \delta \,\sigmatm - 40 \widetilde{\mu}_{+}^{(3)}\right]x^{7}\Biggr\} + \mathcal{O}\left(\frac{\epsilon_\text{tidal}}{c^{6}} \right)\,.
\end{align}
\end{subequations}

\section{Source multipole moments}\label{app:moments}

While the expression for the source mass quadrupole is given in Eq.~\eqref{eq:Iijcirc}, we present here the expressions for the remaining multipole moments for circular orbits, which are used to compute the complete waveform amplitude to 2.5PN order. They read
 \begin{subequations}\label{eq:momentcirc}
	\begin{align}
	& \dI_{ijk}=\tmass\nu r^3
	\Biggl[ n^{\langle i} n^{j} n^{k\rangle} \biggl\{- \delta \left(1 -  \nu \gamma +\left( -\frac{139}{330}-\frac{11923}{660}\nu-\frac{29}{110}\nu^2\right)\gamma^2 \right)
	+ 18 \widetilde{\mu}_{-}^{(2)} \gamma^5 \nn \\
 &	+ \biggl [\delta\, \widetilde{\mu}_{+}^{(2)} \Bigl(- \frac{3}{2}
	+ 48 \nu \Bigr)
	+ \widetilde{\mu}_{-}^{(2)} \Bigl(- \frac{39}{2} - 60 \nu \Bigr)
	- 84 \delta\, \widetilde{\sigma}_{+}^{(2)}+ 84 \widetilde{\sigma}_{-}^{(2)}\biggl]
	\gamma^6 \nn \\
 &  +\left[\delta\,\mutp\left( \frac{19589}{280} +\frac{30101}{220}\nu -\frac{1953}{55}\nu^2 \right)+\mutm\left( \frac{51451}{616}+\frac{106021}{308}\nu -\frac{2819}{11}\nu^2 \right) + \delta\,\sigmatp\left( \frac{328}{3}+\frac{1628}{3}\nu\right) \right. \nn \\
 & \left. + \sigmatm\left(-\frac{616}{3}+164\nu \right) -\frac{15}{\nu}\left( \delta\,\widetilde{\mu}_+^{(3)}+\widetilde{\mu}_-^{(3)}\right)\right]\gamma^7\biggl\}+n^{\langle i} \lambda^{j} \lambda^{k\rangle}
	\biggl\{- \delta \left((1 - 2 \nu ) \gamma  +\left( \frac{571}{165}+\frac{877}{330}\nu-\frac{89}{55}\nu^2\right)\gamma^2 \right)\nonumber \\
	& + \biggl [\delta\,\mutp (- 39 +36\nu) + \mutm (39-42 \nu) - 72 \delta\, \sigmatp+ 72 \sigmatm\biggl]\gamma^6 +\left[\delta\,\mutp\left( \frac{313}{5} -\frac{23819}{110}\nu -\frac{1746}{55}\nu^2 \right) \right. \nn \\
 & \left. +\mutm\left( -\frac{1267}{11}+\frac{8453}{22}\nu -\frac{2105}{11}\nu^2 \right) + \delta\,\sigmatp\left( \frac{340}{3}+\frac{908}{3}\nu\right) + \sigmatm\left(-\frac{532}{3}+196\nu \right)  \right]\gamma^7\biggl\}\Biggl]
	\, ,\\
	&\dI_{ijkl}=\tmass\nu r^4 \Biggl[n^{\langle i} n^{j} n^{k} n^{l\rangle} \biggl\{1 - 3 \nu + \left(\frac{3}{110} -\frac{25}{22}\nu +\frac{69}{22}\nu^2\right)\gamma + \Bigl(18 \mutp	- 18 \delta \,\mutm\Bigr) \gamma^5 \nn\\
 & + \left[\mutp\left(-\frac{2583}{110}-\frac{4182}{55}\nu+\frac{2106}{11}\nu^2\right) +\delta\,\mutm\left(\frac{3957}{110}+\frac{3384}{55}\nu\right)+\sigmatp\left(\frac{864}{5}-\frac{1728}{5}\nu\right)-\frac{864}{5}\delta\,\sigmatm \right]\gamma^6\biggr\}\nn\\
 & + n^{\langle i} n^{j} \lambda^{k} \lambda^{l\rangle} \biggl\{ \left(\frac{78}{55} -\frac{78}{11}\nu +\frac{78}{11}\nu^2\right)\gamma + \left[\mutp\left(\frac{5112}{55}-\frac{15174}{55}\nu+\frac{1404}{11}\nu^2\right) +\delta\,\mutm\left(-\frac{3708}{55}+\frac{2718}{55}\nu\right) \nn \right.\\
 &\left. +\sigmatp\left(\frac{768}{5}-\frac{1536}{5}\nu\right)-\frac{768}{5}\delta\,\sigmatm \right]\gamma^6\biggr\}\Biggr]\, , \\
&\dI_{ijklm}=\tmass\nu r^5 \Biggl[n^{\langle i} n^{j} n^{k} n^{l} n^{m\rangle} \biggl\{-\delta \left(1 - 2 \nu + \left(\frac{2}{39} -\frac{47}{39}\nu +\frac{28}{13}\nu^2\right)\gamma \right) + \Bigl((30-60\nu) \mutm	- 30 \delta \,\mutp\Bigr) \gamma^5 \nn\\
 & + \left[\delta \,\mutp\left(\frac{1361}{26}+\frac{1237}{13}\nu-\frac{2064}{13}\nu^2\right) +\,\mutm\left(-\frac{1697}{26}+\frac{1210}{13}\nu+\frac{2710}{13}\nu^2\right)+\delta \, \sigmatp\left(-\frac{880}{3}+\frac{880}{3}\nu\right) \nn \right. \\
 & \left. +\sigmatm \left(\frac{880}{3}-880\nu \right) \right]\gamma^6\biggr\} + n^{\langle i} n^{j} n^{k} \lambda^{l} \lambda^{m\rangle} \biggl\{ \delta \left(-\frac{70}{39} +\frac{280}{39}\nu -\frac{70}{13}\nu^2\right)\gamma  + \left[\delta \, \mutp\left(-\frac{2020}{13}+\frac{4100}{13}\nu-\frac{1260}{13}\nu^2\right) \nn \right.\\
 & \left. +\mutm\left(\frac{1600}{13}-\frac{5620}{13}\nu+\frac{2420}{13}\nu^2\right)+\delta\, \sigmatp\left(-\frac{800}{3}+\frac{800}{3}\nu\right)+\sigmatm \left(\frac{800}{3}-800\nu \right) \right]\gamma^6\biggr\}\Biggr]\, , \\
&\dI_{ijklmp}=\tmass \nu r^6 n^{\langle i} n^{j} n^{k} n^l n^m n^{p\rangle} \biggl[ 1-5\nu+5\nu^2 +  \left(\mutp (45-135\nu) +\delta\,\mutm(-45+45\nu) \right)\gamma^5 \biggr]\, , \\
&\dI_{ijklmpq}=\tmass \nu r^7 n^{\langle i} n^{j} n^{k} n^l n^m n^p n^{q\rangle} \biggl[ -\delta\Bigl( 1-4\nu+3\nu^2\Bigr) + 63 \left( \mutm(1-4\nu+2\nu^2) +\delta\,\mutp(-1+2\nu) \right)\gamma^5 \biggr]\, , \\
& \dJ_{ij}=\sqrt{G} \,\tmass^{3/2} r^{3/2} \,\ell^{\langle i} n^{j\rangle}\biggl\{- \delta \nu \left[1 + \left(\frac{25}{28} + \frac{3}{14} \nu \right) \gamma + \left(-\frac{17}{63} - \frac{2761}{252} \nu + \frac{19}{84} \nu^2\right) \gamma^2 \right] \nn \\
& + \Bigl(-9  \nu\delta\, \mutp + 9 \nu\, \mutm + 12 \delta\, \sigmatp + 12 \sigmatm\Bigr) \gamma^5 \nn\\ 
&  + \biggl [\delta\,\nu\, \mutp \Bigl(\frac{663}{28} + \frac{117}{7} \nu\Bigr) + \nu\,\mutm \Bigl(- \frac{177}{7} + \frac{477}{14} \nu\Bigr) + \delta\, \sigmatp \Bigl(-10 -  \frac{690}{7} \nu \Bigr) + \sigmatm \Bigl(-10 -  \frac{346}{7} \nu \Bigr)\biggl] \gamma^6 \nn \\
&  + \biggl [\delta\,\nu\, \mutp \Bigl(-\frac{28435}{336} + \frac{1363}{24} \nu-\frac{15}{7}\nu^2\Bigr) + \nu\,\mutm \Bigl(\frac{23323}{336} + \frac{38821}{168} \nu - \frac{16535}{168}\nu^2\Bigr) \nn \\
&  + \delta\, \sigmatp \Bigl(\frac{523}{14} +  \frac{30137}{126} \nu +\frac{38707}{126}\nu^2\Bigr) + \sigmatm \Bigl(\frac{523}{14} +  \frac{4849}{126} \nu + \frac{22433}{42} \nu^2 \Bigr)-60\delta\, \nu \, \widetilde{\mu}_+^{(3)}\biggl] \gamma^7 \biggl\}  \, , \\
	&\dJ_{ijk}=\sqrt{G} \tmass^{3/2} r^{5/2} \nu \Biggl[\ell^{\langle i} n^{j} n^{k\rangle} \biggl\{1 - 3 \nu + \left(\frac{23}{45}-\frac{19}{18}\nu-\frac{7}{9}\nu^2\right)\gamma
	+ \biggl [\mutp \Bigl(21 - 27 \nu \Bigr) - 12 \delta\, \mutm
	+ 64 \sigmatp\biggl] \gamma^5 \nn \\
 & +\left[ \mutp\left(-\frac{1059}{20} +\frac{363}{4}\nu +\frac{183}{2}\nu^2 \right)+\delta\,\mutm\left(\frac{181}{4} -\frac{65}{4}\nu \right)+\sigmatp\left(-48-\frac{976}{3}\nu \right)+\frac{128}{3}\delta\,\sigmatm\right]\gamma^6 \biggl\} \nn\\
 & +\ell^{\langle i} \lambda^{j} \lambda^{k\rangle} \biggl\{ \left(\frac{7}{45}-\frac{7}{9}\nu+\frac{7}{9}\nu^2\right)\gamma +\left[ \mutp\left(\frac{61}{5}-36\nu+21\nu^2\right)+\delta\,\mutm\left(-8+5\nu \right)+\sigmatp\left(\frac{64}{3}-\frac{128}{3}\nu \right) \right. \nn \\
 & \left. -\frac{64}{3}\delta\,\sigmatm\right]\gamma^6 \biggl\}\Biggr]\,,\\
	&\dJ_{ijkl}=\sqrt{G} \tmass^{3/2} r^{7/2} \nu \Biggl[\ell^{\langle i} n^{j} n^{k} n^{l\rangle} \biggl\{-\delta (1 - 2 \nu) + \delta \left(-\frac{7}{22}+\frac{\nu}{44}+\frac{6}{11}\nu^2\right)\gamma
	+ \biggl [\delta\,\mutp \left(-\frac{63}{2} +18 \nu \right)\nn \\
 &  +\delta\, \mutm\left(\frac{45}{2}-45\nu\right) - 60\delta\,\sigmatp+60\sigmatm\biggl] \gamma^5 +\left[ \delta\,\mutp\left(\frac{1935}{22} -\frac{3123}{44}\nu -\frac{993}{11}\nu^2 \right)\right.\nn \\
 & \left. +\mutm\left(-\frac{1695}{22} +\frac{5097}{22}\nu - \frac{387}{22}\nu^2 \right) +\delta\,\sigmatp\left(\frac{1182}{11}+\frac{2646}{11}\nu \right)  +\sigmatm\left(-\frac{1710}{11}+\frac{730}{11}\nu \right)\right]\gamma^6 \biggl\} \nn \\
 & +\ell^{\langle i} n^{j} \lambda^{k} \lambda^{l\rangle} \biggl\{ \delta\left(-\frac{4}{11}+\frac{16}{11}\nu-\frac{12}{11}\nu^2\right)\gamma+ \left[ \delta\,\mutp\left(-\frac{801}{22}+\frac{819}{11}\nu-\frac{324}{11}\nu^2\right) \right. \nn\\
 & \left. +\mutm\left(\frac{585}{22}-\frac{972}{11}\nu +\frac{387}{11}\nu^2 \right)+\delta\,\sigmatp\left(-\frac{780}{11}+\frac{780}{11}\nu \right)+\sigmatm\left(\frac{780}{11}-\frac{2340}{11}\nu \right)\right]\gamma^6 \biggl\}\Biggr]\,,\\
	&\dJ_{ijklm}=\sqrt{G}\tmass^{3/2} r^{9/2} \nu\, \ell^{\langle i} n^{j} n^{k} n^l n^{m\rangle} \biggl\{ 1-5\nu+5\nu^2 + \left[ \mutp\left(45-153\nu+45\nu^2\right)+ \delta\,\mutm\left(-36+36\nu\right) \right. \nn \\
 & \qquad \qquad \left.  +\sigmatp (96-192\nu)-96\delta\, \sigmatm \right]\gamma^5 \biggr\}\,,\\
	&\dJ_{ijklmp}=\sqrt{G}\tmass^{3/2} r^{11/2} \nu \,\ell^{\langle i} n^{j} n^{k} n^l n^m n^{p\rangle} \biggl\{ -\delta\Bigl( 1-4\nu+3\nu^2\Bigr) + \left[ \delta\,\mutp\left(-\frac{123}{2}+141\nu-27\nu^2\right) \right. \nn \\
 & \qquad \qquad \left. +\mutm\left(\frac{105}{2}-210\nu+105\nu^2\right) +140\delta\,\sigmatp (-1+\nu) + \sigmatm(140-240\nu) \right]\gamma^5 \biggr\}\,.
\end{align}
\end{subequations}

\section{Factorized modes}\label{app:EOBmodes}

Here, we provide the expressions for the amplitudes $f_{\ell m}$ in the factorized modes, with $\ell \leq 7$ and $\lvert m \rvert \leq \ell $, up to 2.5PN order. The (2,2) mode can be found in Eq.~\eqref{eq:f22} and the rest reads
\begin{subequations}\label{eq:flmvalues}
\begin{align}
f^\text{tidal}_{21}&= \frac{1}{\delta\,\nu} \Biggl\{6 \delta\,\nu\,\mutp -9 \nu\, \mutm -12\delta\,\sigmatp-12\sigmatm +\left[ \delta\,\mutp\left(-\frac{181}{28}\nu -\frac{71}{7}\nu^2 \right) +\mutm\left( \frac{239}{28}\nu -\frac{379}{14}\nu^2 \right) \nn \right. \\
& \quad \left. +\delta\,\sigmatp\left(-14 +\frac{354}{7}\nu \right) +\sigmatm\left(-14 +\frac{458}{7}\nu \right) \right]x + \left[\delta\,\mutp\left(-\frac{1195}{48}\nu -\frac{3267}{56}\nu^2 +\frac{55}{14}\nu^3 \right) \nn \right. \\
& \quad \left. +\mutm\left(-\frac{11083}{336}\nu -\frac{7655}{84}\nu^2 +\frac{11567}{168}\nu^3 \right)  +\delta\,\sigmatp\left(-\frac{355}{14} +\frac{25051}{126}\nu-\frac{3025}{18}\nu^2 \right) \nn \right. \\
& \quad \left.+\sigmatm\left(-\frac{355}{14} +\frac{28451}{126}\nu-\frac{1795}{6}\nu^2\right) + 40\delta\,\widetilde{\mu}_+^{(3)}\nu\right]x^2 \Biggr\}+ \calO\left( \frac{\etidal}{c^6}\right)\,,\\
f^\text{tidal}_{33}&= \frac{18}{\delta}\Biggl\{ \delta\,\mutp - \mutm+  \left[\delta\,\mutp\left( -\frac{11}{3} +\frac{5}{6}\nu \right) + \mutm\left(\frac{17}{6} -\frac{20}{3}\nu \right) +6\delta\,\sigmatp -\frac{2}{3}\sigmatm \right]x \nn \\
& \quad+\left[ \delta\,\mutp \left( -\frac{17459}{6930} -\frac{387391}{27720}\nu- \frac{391}{990}\nu^2 \right) + \mutm\left( -\frac{1417}{1386}+\frac{32891}{5544}\nu-\frac{358}{99}\nu^2\right)+ \delta\,\sigmatp \left( -\frac{1}{9}-\frac{74}{9}\nu \right) \right. \nn \\
& \quad \quad \left. + \sigmatm\left(\frac{31}{3} -42\nu \right) + \delta\,\widetilde{\mu}_+^{(3)} \left( \frac{5}{6\nu} + \frac{20}{3}\right) +\frac{5}{6\nu}\widetilde{\mu}_-^{(3)}  \right]x^2\Biggr\}+ \calO\left( \frac{\etidal}{c^6}\right)\,,\\
f^\text{tidal}_{32}&= \frac{12}{1-3\nu}\Biggl\{ \mutp(2-3\nu) -\delta\,\mutm +\frac{16}{3} \sigmatp+\left[ \mutp\left( -\frac{697}{180} + \frac{635}{72}\nu +\frac{26}{9}\nu^2 \right) + \delta\,\mutm \left( \frac{55}{12} -\frac{101}{24}\nu\right) \nn \right.\\
& \quad \left. +\sigmatp\left( -\frac{28}{9}-\frac{188}{9}\nu\right) +\frac{16}{3} \delta\,\sigmatm \right]x \Biggr\}+ \calO\left( \frac{\etidal}{c^4}\right) \,,\\
f^\text{tidal}_{31}&= \frac{18}{\delta}\Biggl\{ \delta\,\mutp - \mutm+  \left[\delta\,\mutp\left( \frac{1}{9} -\frac{65}{18}\nu \right) + \mutm\left(-\frac{1}{18} -\frac{32}{9}\nu \right) +\frac{34}{3}\delta\,\sigmatp -6\sigmatm \right]x \nn \\
& \quad+\left[ \delta\,\mutp \left( \frac{18065}{4158} -\frac{45659}{5544}\nu+ \frac{199}{66}\nu^2 \right) + \mutm\left( -\frac{1039}{1386}-\frac{330343}{16632}\nu+\frac{2204}{99}\nu^2\right)+ \delta\,\sigmatp \left( \frac{1207}{81}-\frac{3682}{81}\nu \right) \right. \nn \\
& \quad \quad \left. + \sigmatm\left(\frac{389}{81} -\frac{1382}{27}\nu \right) + \delta\,\widetilde{\mu}_+^{(3)} \left( \frac{5}{6\nu} + \frac{20}{3}\right) +\frac{5}{6\nu}\widetilde{\mu}_-^{(3)}  \right]x^2\Biggr\}+ \calO\left( \frac{\etidal}{c^6}\right)\,,\\
f^\text{tidal}_{44}&= \frac{6}{1-3\nu}\Biggl\{ \mutp(7-12\nu) -3\delta\,\mutm +\left[ \mutp\left( -\frac{17321}{660} + \frac{12596}{165}\nu -\frac{335}{22}\nu^2 \right) + \delta\,\mutm \left( \frac{13903}{660} -\frac{1209}{55}\nu\right) \nn \right.\\
& \quad \left. +\sigmatp\left( \frac{368}{15}-\frac{352}{5}\nu\right) -\frac{16}{5} \delta\,\sigmatm \right]x \Biggr\}+ \calO\left( \frac{\etidal}{c^4}\right) \,,\\
f^\text{tidal}_{43}&= \frac{60}{\delta(1-2\nu)}\Biggl\{ \delta\,\mutp\left(\frac{27}{40}-\frac{3}{5}\nu\right) +\mutm\left( -\frac{3}{8} +\frac{3}{4}\nu \right) +\delta\,\sigmatp -\sigmatm +\left[ \delta\,\mutp\left( -\frac{39}{16} + \frac{184}{55}\nu +\frac{2}{11}\nu^2 \right) \nn \right.\\
& \quad \left. + \mutm \left( \frac{381}{176} -\frac{424}{55}\nu +\frac{1707}{440}\nu^2\right)  +\delta\,\sigmatp\left( -\frac{327}{110}-\frac{1153}{330}\nu\right) +\sigmatm\left( \frac{83}{22} -\frac{263}{66}\nu \right)  \right]x \Biggr\}+ \calO\left( \frac{\etidal}{c^4}\right) \,,\\
f^\text{tidal}_{42}&= \frac{6}{1-3\nu}\Biggl\{ \mutp(7-12\nu) -3\delta\,\mutm +\left[ \mutp\left( -\frac{4289}{660} + \frac{1499}{165}\nu +\frac{601}{22}\nu^2 \right) + \delta\,\mutm \left( \frac{6487}{660} -\frac{756}{55}\nu\right) \nn \right.\\
& \quad \left. +\sigmatp\left( \frac{752}{15}-\frac{608}{5}\nu\right) -\frac{144}{5} \delta\,\sigmatm \right]x \Biggr\} + \calO\left( \frac{\etidal}{c^4}\right)\,,\\
f^\text{tidal}_{41}&= \frac{60}{\delta(1-2\nu)}\Biggl\{ \delta\,\mutp\left(\frac{27}{40}-\frac{3}{5}\nu\right) +\mutm\left( -\frac{3}{8} +\frac{3}{4}\nu \right) +\delta\,\sigmatp -\sigmatm +\left[ \delta\,\mutp\left( -\frac{77}{48} + \frac{263}{165}\nu +\frac{10}{11}\nu^2 \right) \nn \right.\\
& \quad \left. + \mutm \left( \frac{277}{176} -\frac{316}{55}\nu +\frac{1363}{440}\nu^2\right)  +\delta\,\sigmatp\left( -\frac{461}{330}-\frac{1673}{330}\nu\right) + \sigmatm \left( \frac{145}{66} +\frac{49}{66}\nu \right)  \right]x \Biggr\} + \calO\left( \frac{\etidal}{c^4}\right)\,,\\
f^\text{tidal}_{55}&= \frac{30}{\delta(1-2\nu)} \Biggl\{ \delta\,\mutp(2-2\nu) +\mutm(-1+2\nu) +\left[ \delta\,\mutp \left( -\frac{4109}{390} + \frac{2477}{130}\nu-\frac{229}{65}\nu^2 \right) \nn \right. \\
& \quad \left. + \mutm \left(\frac{511}{65} - \frac{2263}{78}\nu+\frac{207}{13}\nu^2\right) +\delta\,\sigmatp\left(\frac{56}{9} -\frac{104}{9}\nu \right) + \sigmatm\left( -\frac{8}{9} +\frac{8}{3}\nu \right)\right]x \Biggr\}+ \calO\left( \frac{\etidal}{c^4}\right) \,,\\
f^\text{tidal}_{54}&= f^\text{tidal}_{52} = \frac{96}{1-5\nu+5\nu^2} \left[ \mutp \left(\frac{5}{8}-\frac{19}{8}\nu+\frac{5}{4}\nu^2\right) +\delta\,\mutm \left(-\frac{3}{8}+\frac{3}{8}\nu\right)+\sigmatp(1-2\nu) -\delta\,\sigmatm\right]+ \calO\left( \frac{\etidal}{c^2}\right)\,,\\
f^\text{tidal}_{53}&= \frac{30}{\delta(1-2\nu)} \Biggl\{ \delta\,\mutp(2-2\nu) +\mutm(-1+2\nu) +\left[ \delta\,\mutp \left( -\frac{409}{78} + \frac{2359}{390}\nu+\frac{163}{65}\nu^2 \right) \nn \right. \\
& \quad \left. + \mutm \left(\frac{893}{195} - \frac{2273}{130}\nu+\frac{2137}{195}\nu^2\right) +\delta\,\sigmatp\left(\frac{40}{3} -\frac{56}{3}\nu \right) + \sigmatm\left( -8 +24\nu \right)\right]x \Biggr\}+ \calO\left( \frac{\etidal}{c^4}\right) \,,\\
f^\text{tidal}_{51}&= \frac{30}{\delta(1-2\nu)} \Biggl\{ \delta\,\mutp(2-2\nu) +\mutm(-1+2\nu) +\left[ \delta\,\mutp \left( -\frac{1013}{390} - \frac{59}{130}\nu+\frac{359}{65}\nu^2 \right) \nn \right. \\
& \quad \left. + \mutm \left(\frac{191}{65} - \frac{4571}{390}\nu+\frac{551}{65}\nu^2\right) +\delta\,\sigmatp\left(\frac{152}{9} -\frac{200}{9}\nu \right) + \sigmatm\left( -\frac{104}{9} +\frac{104}{3}\nu \right)\right]x \Biggr\} + \calO\left( \frac{\etidal}{c^4}\right)\,,\\
f^\text{tidal}_{66}&=f^\text{tidal}_{64} = f^\text{tidal}_{62} = \frac{9}{1-5\nu+5\nu^2} \left[ \mutp (9-35\nu+20\nu^2) +\delta\,\mutm (-5+5\nu)\right]+ \calO\left( \frac{\etidal}{c^2}\right) \,,\\
f^\text{tidal}_{65}&= f^\text{tidal}_{63} = f^\text{tidal}_{61}=\frac{140}{\delta(1-\nu)(1-3\nu)}\biggl[\delta\,\mutp \left(\frac{33}{56}-\frac{45}{28}\nu+\frac{9}{14}\nu^2\right)+\mutm\left(-\frac{3}{8}+\frac{3}{2}\nu-\frac{3}{4}\nu^2\right) \nn \\
& \qquad \qquad \qquad \qquad \qquad \qquad \qquad \qquad +\delta\,\sigmatp (1-\nu) +\sigmatm(-1+3\nu)\biggr]+ \calO\left( \frac{\etidal}{c^2}\right)\,,\\
f^\text{tidal}_{77}&= f^\text{tidal}_{75} =f^\text{tidal}_{73}=f^\text{tidal}_{71} = \frac{21}{\delta(1-\nu)(1-3\nu)} \biggl[\delta\,\mutp (5-14\nu+6\nu^2)+\mutm(-3+12\nu-6\nu^2) \biggr]+ \calO\left( \frac{\etidal}{c^2}\right)\,.
\end{align}
\end{subequations}
Finally the expressions for the $\rho_{\ell m} = (f_{\ell m})^{1/\ell}$ are given by
\begin{subequations}\label{eq:rholm}
\begin{align}
\rho^\text{tidal}_{21}&= \frac{1}{\delta\,\nu} \Biggl\{3 \delta\,\nu\,\mutp -\frac{9}{2} \nu\, \mutm -6\delta\,\sigmatp-6\sigmatm +\left[ \delta\,\mutp\left(-\frac{\nu}{14} -\frac{165}{28}\nu^2 \right) +\mutm\left( -\frac{53}{112}\nu -\frac{689}{56}\nu^2 \right) \nn \right. \\
& \quad \left. +\delta\,\sigmatp\left(-\frac{373}{28} +\frac{377}{14}\nu \right) +\sigmatm\left(-\frac{373}{28} +\frac{481}{14}\nu \right) \right]x + \left[\delta\,\mutp\left(-\frac{188627}{18816}\nu -\frac{25889}{784}\nu^2 +\frac{4993}{1568}\nu^3 \right) \nn \right. \\
& \quad \left. +\mutm\left(-\frac{111491}{5376}\nu -\frac{27733}{448}\nu^2 +\frac{51589}{1344}\nu^3 \right)  +\delta\,\sigmatp\left(-\frac{298331}{9408} +\frac{894373}{7056}\nu-\frac{91339}{1008}\nu^2 \right) \nn \right. \\
& \quad \left.+\sigmatm\left(-\frac{298331}{9408} +\frac{1044797}{7056}\nu-\frac{372095}{2352}\nu^2\right) + 20\delta\,\widetilde{\mu}_+^{(3)} \nu\right]x^2 \Biggr\}+ \calO\left( \frac{\etidal}{c^6}\right)\,,\\
\rho^\text{tidal}_{33}&= \frac{6}{\delta}\Biggl\{ \delta\,\mutp - \mutm+  \left[\delta\,\mutp\left( -\frac{4}{3} -\frac{\nu}{2} \right) + \mutm\left(\frac{1}{2} -\frac{16}{3}\nu \right) +6\delta\,\sigmatp -\frac{2}{3}\sigmatm \right]x \nn \\
& \quad+\left[ \delta\,\mutp\left( -\frac{4987}{1386}-\frac{44623}{5544}\nu-\frac{71}{66}\nu^2\right)  + \mutm\left( -\frac{6542}{3465} -\frac{346321}{27720}\nu+ \frac{799}{165}\nu^2 \right)+ \delta\,\sigmatp \left( \frac{125}{9}-\frac{146}{9}\nu \right) \right. \nn \\
& \quad \quad \left. + \sigmatm\left(\frac{79}{9} -\frac{370}{9}\nu \right) + \delta\,\widetilde{\mu}_+^{(3)} \left( \frac{5}{6\nu} + \frac{20}{3}\right) +\frac{5}{6\nu}\widetilde{\mu}_-^{(3)}  \right]x^2\Biggr\}+ \calO\left( \frac{\etidal}{c^6}\right)\,,\\
\rho^\text{tidal}_{32}&= \frac{4}{(1-3\nu)^2}\Biggl\{ (1-3\nu)\biggl(\mutp(2-3\nu) -\delta\,\mutm +\frac{16}{3} \sigmatp\biggr)+\left[ \mutp\left( \frac{533}{540} - \frac{3641}{1080}\nu +\frac{1285}{216}\nu^2 - \frac{142}{9}\nu^3 \right)  \nn \right.\\
& \quad \left. + \delta\,\mutm \left( \frac{1163}{540} -\frac{2095}{216}\nu+\frac{2215}{216}\nu^2\right) +\sigmatp\left( \frac{3988}{405}-\frac{4504}{81}\nu +\frac{6100}{81}\nu^2\right) +\frac{16}{3} (1-3\nu) \delta\,\sigmatm \right]x \Biggr\}+ \calO\left( \frac{\etidal}{c^4}\right) \,,\\
\rho^\text{tidal}_{31}&= \frac{6}{\delta}\Biggl\{ \delta\,\mutp - \mutm+  \left[\delta\,\mutp\left( \frac{14}{9} -\frac{19}{6}\nu \right) + \mutm\left(-\frac{3}{2} -4\nu \right) +\frac{34}{3}\delta\,\sigmatp -6\sigmatm \right]x \nn \\
& \quad+\left[ \delta\,\mutp\left( \frac{25121}{4158}-\frac{526319}{49896}\nu+\frac{4435}{1782}\nu^2\right)  + \mutm\left( -\frac{14759}{6237} -\frac{1390925}{49896}\nu+ \frac{17467}{891}\nu^2 \right) \right. \nn \\
& \quad \quad \left. + \delta\,\sigmatp \left( \frac{2533}{81}-\frac{3274}{81}\nu \right) + \sigmatm\left(-\frac{313}{81} -\frac{1454}{27}\nu \right) + \delta\,\widetilde{\mu}_+^{(3)} \left( \frac{5}{6\nu} + \frac{20}{3}\right) +\frac{5}{6\nu}\widetilde{\mu}_-^{(3)}  \right]x^2\Biggr\}+ \calO\left( \frac{\etidal}{c^6}\right)\,,\\
\rho^\text{tidal}_{44}&= \frac{1}{(1-3\nu)^2}\Biggl\{ (1-3\nu)\left(\mutp\left(\frac{21}{2}-18\nu\right) -\frac{9}{2}\delta\,\mutm\right) +\left[ \mutp\left( -\frac{17}{20} + \frac{53}{2}\nu -\frac{55959}{880}\nu^2 -\frac{855}{22}\nu^3 \right)  \nn \right.\\
& \quad \left. + \delta\,\mutm \left( \frac{166}{11} -\frac{14901}{220}\nu+\frac{63423}{880}\nu^2\right) +\sigmatp\left( \frac{184}{5}-216\nu + \frac{1584}{5}\nu^2\right) -\frac{24}{5}(1-3\nu) \delta\,\sigmatm \right]x \Biggr\}+ \calO\left( \frac{\etidal}{c^4}\right) \,,\\
\rho^\text{tidal}_{43}&= \frac{15}{\delta(1-2\nu)^2}\Biggl\{ \delta\,\mutp\left(\frac{27}{40}-\frac{39}{20}\nu+\frac{6}{5}\nu^2\right) +\mutm\left( -\frac{3}{8} +\frac{3}{2}\nu -\frac{3}{2}\nu^2 \right) +(1-2\nu)\bigl(\delta\,\sigmatp  -\sigmatm \bigr)  \nn \\
& \quad +\left[ \delta\,\mutp\left( \frac{411}{3520} - \frac{2419}{7040}\nu +\frac{163}{176}\nu^2 -2\nu^3 \right)+ \mutm \left( \frac{525}{704} -\frac{40157}{7040}\nu +\frac{39713}{3520}\nu^2-\frac{1257}{220}\nu^3\right) \right. \nn \\
& \quad \left. +\delta\,\sigmatp\left( \frac{357}{440}-\frac{18143}{2640}\nu+\frac{1603}{165}\nu^2\right) +\sigmatm\left( -\frac{1}{88} -\frac{1165}{528}\nu +\frac{173}{33}\nu^2\right)  \right]x \Biggr\}+ \calO\left( \frac{\etidal}{c^4}\right) \,,\\
\rho^\text{tidal}_{42}&= \frac{1}{(1-3\nu)^2}\Biggl\{ (1-3\nu)\left(\mutp\left(\frac{21}{2}-18\nu\right) -\frac{9}{2}\delta\,\mutm\right) +\left[ \mutp\left( \frac{88}{5} - \frac{353}{4}\nu +\frac{133149}{880}\nu^2 -\frac{2961}{22}\nu^3 \right)  \nn \right.\\
& \quad \left. + \delta\,\mutm \left( \frac{133}{44} -\frac{1581}{55}\nu+\frac{51867}{880}\nu^2\right) +\sigmatp\left( \frac{376}{5}-408\nu + \frac{2736}{5}\nu^2\right) -\frac{216}{5}(1-3\nu) \delta\,\sigmatm \right]x \Biggr\}+ \calO\left( \frac{\etidal}{c^4}\right) \,,\\
\rho^\text{tidal}_{41}&= \frac{15}{\delta(1-2\nu)^2}\Biggl\{ \delta\,\mutp\left(\frac{27}{40}-\frac{39}{20}\nu+\frac{6}{5}\nu^2\right) +\mutm\left( -\frac{3}{8} +\frac{3}{2}\nu -\frac{3}{2}\nu^2 \right) +(1-2\nu)\bigl(\delta\,\sigmatp  -\sigmatm \bigr)  \nn \\
& \quad +\left[ \delta\,\mutp\left( \frac{7441}{10560} - \frac{3607}{1408}\nu +\frac{1873}{528}\nu^2 -\frac{14}{5}\nu^3 \right)+ \mutm \left( \frac{205}{704} -\frac{23773}{7040}\nu +\frac{28417}{3520}\nu^2-\frac{1093}{220}\nu^3\right) \right. \nn \\
& \quad \left. +\delta\,\sigmatp\left( \frac{2671}{1320}-\frac{26783}{2640}\nu+\frac{1943}{165}\nu^2\right) +\sigmatm\left( -\frac{323}{264} +\frac{2227}{528}\nu -\frac{103}{33}\nu^2\right)  \right]x \Biggr\}+ \calO\left( \frac{\etidal}{c^4}\right) \,,\\
\rho^\text{tidal}_{55}&= \frac{6}{\delta(1-2\nu)^2} \Biggl\{ (1-2\nu) \bigl( \delta\,\mutp(2-2\nu) +\mutm(-1+2\nu) \bigr) +\left[ \delta\,\mutp \left( -\frac{71}{130} + \frac{1369}{390}\nu-\frac{878}{195}\nu^2 -\frac{674}{195}\nu^3 \right) \nn \right. \\
& \quad \left. + \mutm \left(\frac{43}{15} - \frac{643}{30}\nu+\frac{8204}{195}\nu^2-\frac{4162}{195}\nu^3\right) +\delta\,\sigmatp\left(\frac{56}{9} -24\nu +\frac{208}{9}\nu^2\right) + \sigmatm\left( -\frac{8}{9} +\frac{40}{9}\nu-\frac{16}{3}\nu^2 \right)\right]x \Biggr\} \nn \\
& \quad+ \calO\left( \frac{\etidal}{c^4}\right) \,,\\
\rho^\text{tidal}_{54}&= \rho^\text{tidal}_{52} = \frac{96}{5(1-5\nu+5\nu^2)} \left[ \mutp \left(\frac{5}{8}-\frac{19}{8}\nu+\frac{5}{4}\nu^2\right) +\delta\,\mutm \left(-\frac{3}{8}+\frac{3}{8}\nu\right)+\sigmatp(1-2\nu) -\delta\,\sigmatm\right]+ \calO\left( \frac{\etidal}{c^2}\right)\,,\\
\rho^\text{tidal}_{53}&= \frac{6}{\delta(1-2\nu)^2} \Biggl\{ (1-2\nu) \bigl( \delta\,\mutp(2-2\nu) +\mutm(-1+2\nu) \bigr) +\left[ \delta\,\mutp \left( \frac{191}{78} - \frac{1117}{130}\nu+\frac{2234}{195}\nu^2 -\frac{1682}{195}\nu^3 \right) \nn \right. \\
& \quad \left. + \mutm \left(\frac{11}{15} - \frac{307}{30}\nu+\frac{5204}{195}\nu^2-\frac{238}{13}\nu^3\right) +\delta\,\sigmatp\left(\frac{40}{3} -\frac{136}{3}\nu +\frac{112}{3}\nu^2\right) + \sigmatm\left( -8 +40\nu-48\nu^2 \right)\right]x \Biggr\} \nn \\
& \quad+ \calO\left( \frac{\etidal}{c^4}\right) \,,\\
\rho^\text{tidal}_{51}&= \frac{6}{\delta(1-2\nu)^2} \Biggl\{ (1-2\nu) \bigl( \delta\,\mutp(2-2\nu) +\mutm(-1+2\nu) \bigr) +\left[ \delta\,\mutp \left( \frac{513}{130} - \frac{5711}{390}\nu+\frac{758}{39}\nu^2 -\frac{2186}{195}\nu^3 \right) \nn \right. \\
& \quad \left. + \mutm \left(-\frac{1}{3} - \frac{139}{30}\nu+\frac{3704}{195}\nu^2-\frac{3274}{195}\nu^3\right) +\delta\,\sigmatp\left(\frac{152}{9} -56\nu +\frac{400}{9}\nu^2\right) \right. \nn \\
& \quad \left. + \sigmatm\left( -\frac{104}{9} +\frac{520}{9}\nu-\frac{208}{3}\nu^2 \right)\right]x \Biggr\} + \calO\left( \frac{\etidal}{c^4}\right) \,,\\
\rho^\text{tidal}_{66}&=\rho^\text{tidal}_{64} = \rho^\text{tidal}_{62} = \frac{3}{2(1-5\nu+5\nu^2)} \left[ \mutp (9-35\nu+20\nu^2) +\delta\,\mutm (-5+5\nu)\right]+ \calO\left( \frac{\etidal}{c^2}\right) \,,\\
\rho^\text{tidal}_{65}&= \rho^\text{tidal}_{63} = \rho^\text{tidal}_{61}=\frac{70}{3\delta(1-\nu)(1-3\nu)}\biggl[\delta\,\mutp \left(\frac{33}{56}-\frac{45}{28}\nu+\frac{9}{14}\nu^2\right)+\mutm\left(-\frac{3}{8}+\frac{3}{2}\nu-\frac{3}{4}\nu^2\right) \nn \\
& \qquad \qquad \qquad \qquad \qquad \qquad \qquad \qquad +\delta\,\sigmatp (1-\nu) +\sigmatm(-1+3\nu)\biggr]+ \calO\left( \frac{\etidal}{c^2}\right)\,,\\
\rho^\text{tidal}_{77}&= \rho^\text{tidal}_{75} =\rho^\text{tidal}_{73}=\rho^\text{tidal}_{71} = \frac{3}{\delta(1-\nu)(1-3\nu)} \biggl[\delta\,\mutp (5-14\nu+6\nu^2)+\mutm(-3+12\nu-6\nu^2) \biggr]+ \calO\left( \frac{\etidal}{c^2}\right)\,.
\end{align}
\end{subequations}

\bibliography{RefList_TidalModes}

\end{document}